\documentclass[a4paper,12pt]{article}

\usepackage{slashed}
\usepackage[affil-sl,auth-sc]{authblk}
\usepackage{amssymb,amsmath,amsbsy}
\usepackage{xfrac}
\usepackage{mathrsfs}
\usepackage{mathbbol}
\usepackage{bm}
\usepackage{appendix}
\usepackage{hyperref}
\usepackage{cancel}
\usepackage[top=1.2in, bottom=1.2in,left=0.9in,includefoot]{geometry}               
\usepackage{graphicx}
\usepackage{cite}
\usepackage{float}
\usepackage{caption}
\usepackage{wasysym}
\usepackage{amsthm}
\newcommand{\bea}{\begin{eqnarray}}
\newcommand{\eea}{\end{eqnarray}}

\usepackage{color}

\definecolor{orange}{rgb}{0.9,0.2,0}
\definecolor{brown}{rgb}{0.7,0.3,0.2}
\definecolor{fuxia}{rgb}{1,0,1}
\definecolor{skyblue}{rgb}{0,0.1,0.9}
\definecolor{violetred}{rgb}{0.8,0.13,0.56}
\definecolor{deeppink}{rgb}{1.00,0.08,0.5}
\definecolor{pink}{rgb}{1.00,0.75,0.80}
\definecolor{orchid}{rgb}{0.85,0.44,0.84}
\definecolor{lightpink}{rgb}{1.00,0.71,0.76}
\definecolor{bluish}{rgb}{0,0.6,0.8}
\numberwithin{equation}{section}
\title{\bf Diphoton Signal of light pseudoscalar in NMSSM at the LHC}
\author{\small Monoranjan Guchait \thanks{guchait@tifr.res.in}~}
\author{~\small Jacky Kumar\thanks{jka@tifr.res.in}}
\affil{\small Department of High Energy Physics, \\
Tata Institute of Fundamental Research, \\
 Homi Bhabha Road, Mumbai-400005, India}

\def \PMET{E{\!\!\!/}_T}
\def\invfb{fb^{-1}}

\def\t1 {\widetilde {t_1}}
\def\N1{\widetilde \chi_1^0}
\def\N2{\widetilde \chi_2^0}
\def\N3{\widetilde \chi_3^0}
\def\N0{\widetilde \chi^0}
\def\C1{\widetilde \chi_1^{\pm}}
\def\mst1 {m_{\t1}}
\def\br {\begin{eqnarray}}
\def\er {\end{eqnarray}}
\def\bul{\bullet}
\newcommand{\decay}[2]{
\begin{picture}(25,20)(-3,3)
\put(0,-20){\line(0,1){15}}
\put(0,-20){\vector(1,0){15}}
\put(0,0){\makebox(0,0)[lb]{\ensuremath{#1}}}
\put(25,-20){\makebox(0,0)[lc]{\ensuremath{#2}}}
\end{picture}}
\renewcommand{\baselinestretch}{1.2}

\begin{document}
\maketitle
%%%%%%%%%%%%%%%%%%%%%%%%%%%%%%%%%%%%%%%%%%%%%%%%%%%%%%%%%%%%%%%%%%%%%%%%%%%
\begin{abstract}
We explore the detection possibility of light pseudoscalar Higgs boson 
in the next-to-minimal supersymmetric Standard Model(NMSSM) at the LHC with the
center of mass energy, $\sqrt{S}=13$~TeV. We focus on the parameter space 
which provides one of the Higgs boson as the SM-like with a mass of 125 GeV 
and some of the non-SM-like Higgs bosons can be light having suppressed 
couplings with fermions and gauge bosons due to their singlet nature.
It is observed that for certain region of model parameter space,
the singlet like light pseudoscalar can decay to di-photon($\gamma\gamma$) 
channel with a substantial branching ratio. In this study, we consider this 
di-photon signal of light pseudoscalar Higgs boson producing it through the 
chargino-neutralino production and the subsequent decay of neutralino. 
We consider signal consisting of two photons plus missing energy along with a 
lepton from the chargino decay. Performing a detailed simulation of the signal
and backgrounds including detector effects, we present results for a 
few benchmark points corresponding to the pseudoscalar Higgs 
boson mass in the range 60 -100 GeV.
Our studies indicate that some of the benchmark points in the parameter space 
can be probed with a reasonable significance for 100~fb$^{-1}$ 
integrated luminosity. We also conclude that exploiting this channel it is
possible to distinguish the NMSSM from the other supersymmetric models. 
\end{abstract}
\vskip .5 true cm
%\pacs{73.21.Hb, 73.21.La, 73.50.Bk}
%\maketitle
\newpage

\section{Introduction}
%%%%%%%%%%%%%%%%%%%%%%%%%%%%%%%%%%%%%%%%%%%%%%%%%%
In spite of the absence of any signal of
superpartners at the LHC, still supersymmetry(SUSY) remains
one of the best possible option for the physics beyond 
standard model(BSM). Looking for its signal is a very high priority task
in the next phase of LHC experiments.
The SUSY models provide a solution for hierarchy
problem, unify gauge couplings at a certain high energy scale
and in addition, offers a dark matter candidate which is absent in
the standard model(SM).
In order to interpret the recently discovered Higgs particle($\rm H_{SM}$)
of mass $\sim$ 125 GeV at the LHC \cite{Aad:2012tfa,Chatrchyan:2012xdj} 
in the framework of the
minimal supersymmetric standard model(MSSM),
one requires a certain kind of parameter space, in particular for the
squark sector of the third generation~\cite{Hall:2011aa,Arbey:2011ab}.
For instance, the lightest Higgs boson
of mass $\sim$ 125~GeV in the MSSM can be obtained 
either by pushing up the lighter top squark mass to a larger value or assuming
a maximal mixing in the top squark sector.
Moreover, $\mu$ term in the superpotential,
$\rm \mu H_u H_d$ is a another potential source of problem,
where $H_u$ and $ H_d$ are the two Higgs doublets require
to generate the up and down type of fermion masses.
The value of $\mu$ is expected to be around the 
electroweak (EW) scale $\sim {\cal O}$(100~GeV),
but, nothing constrain it not to accept large value, in fact, 
it can go far above the EW scale,
which is known as the $\mu$-problem \cite{Kim:1983dt}.
In the framework of
the Next-to-minimal supersymmetric model (NMSSM) these issues can be
addressed more naturally \cite{Fayet:1974pd,Ellis:1988er,Drees:1988fc}.
The NMSSM contains an extra Higgs singlet field($S$),
in addition to the two Higgs doublets $\rm H_u, H_d$ like the MSSM and, 
the superpotential reads as,
\br
{  \rm W_{NMSSM}} = { \rm  W_{MSSM}} + \lambda  S  {H_u}
{H_d} + \frac{1}{3}\kappa S^3,
\label{eq:poteq}
\er
where $\lambda$ and $\kappa$ are the dimensionless couplings
and $\rm W_{MSSM}$ is the part of the superpotential in the 
MSSM, except the $\mu$ term. After the electroweak symmetry breaking, the 
vacuum expectation value(VEV) of 
the singlet field (S) $v_s$,
generates the $\mu$ term dynamically, i.e  
$\mu_{eff} = \lambda v_s$.
The Higgs sector of the NMSSM contains
three neutral CP even($\rm H_1,H_2,H_3$; $\rm m_{H_1}< m_{H_2} < m_{H_3}$) 
and two CP odd neutral pseudoscalars($\rm A_1,A_2$; $\rm m_{A_1} < m_{A_2}$)
plus charged Higgs boson ($H^\pm$) states
(for details, see the review of Ref.~\cite{Ellwanger:2009dp} 
and Ref.~\cite{Miller:2003ay}).
The states of the physical neutral Higgs bosons are composed of both the
singlet and the doublet fields. Interestingly, one of the
CP even neutral Higgs boson can be interpreted as the recently found 
SM-like Higgs boson and it remains valid for a wide range of 
model parameters~\cite{Kang:2012sy,Cao:2013gba,Vasquez:2012hn,King:2012is,Heinemeyer:2011aa,Domingo:2015eea} 
and, unlike the MSSM, it does 
not require much fine tuning of the model parameters.
It can be attributed to the
mixing of the singlet Higgs field with the doublets via
$\lambda S {H_u} {H_d}$ term. As a consequence, this interaction, in turn 
lifts the tree level Higgs boson mass substantially and then 
further contribution due to the radiative
correction enable to achieve the required 
Higgs boson mass of $\sim$ 125 GeV~\cite{Heinemeyer:2011aa,Domingo:2015eea}.
Naturally, with the discovery of the Higgs 
boson \cite{Aad:2012tfa,Chatrchyan:2012xdj},
the NMSSM has drawn a lot attention, in general, to study in more details   
the Higgs sector and the corresponding phenomenology at the LHC with a great 
interest~\cite{Djouadi:2008uw,King:2012is,Vasquez:2012hn,
King:2012tr,Christensen:2013dra,Cao:2013gba,Kumar:2016vhm}.
Previous studies showed that in the NMSSM framework,
the scenario of very light Higgs bosons
($<$125~GeV) exist, while one of the CP even neutral Higgs boson SM like
\cite{Guchait:2015owa,Gunion:2012zd,King:2012is,Vasquez:2012hn,
Ellwanger:2012ke,Badziak:2013bda}.
Notably, these light Higgs
bosons are non-SM like and dominantly singlet in nature and, hence
not excluded by any past experiments due to the suppression of
their production in colliders. Needless to say, in the present context
of continuing Higgs studies in the LHC experiments, 
it is one of the priority to search for these light non SM-like 
Higgs bosons.
 
Already, in Run 1 experiments at the LHC, extensive searches were 
carried out for the lightest CP odd
Higgs boson($A_1$) either producing it directly or via the decay of the
SM-like Higgs boson, $\rm H_{SM} \to\rm A_1 A_1$. 
The CMS experiment performed searches 
through direct production of $ A_1$ and decaying to a pair of 
muons\cite{Chatrchyan:2012am}  
and taus~\cite{Khachatryan:2015baw} for the mass ranges 5.5 - 14 GeV and 
25-80 GeV respectively and, also looked for it in 
the SM Higgs decay in 4$\tau$ final states~\cite{Khachatryan:2015baw}.   
The ATLAS collaboration published results for 
$A_1$ searches, $\rm H_{SM} \to A_1 A_1 \to \mu\mu \tau\tau$ 
decays with a mass 
range 3.7 - 50~GeV~\cite{Aad:2015oqa} and also in 
four photon final states corresponding to
the mass range 10 - 62 GeV~\cite{Aad:2015bua}.
From the non observation of any signal in all those searches, 
the exclusion of cross sections folded with branching ratios(BR) 
for a given channel are presented for the mass range $\sim 5-60$ of $A_1$. 

On the phenomenological side, after the discovery of the Higgs boson at the
LHC, detection prospects of all Higgs bosons in the NMSSM are 
revisited~\cite{Ellwanger:2004gz,Ellwanger:2011sk,King:2014xwa}.  
Nonetheless, it is more appealing to explore the 
detection possibility of the light non SM-like Higgs bosons in various 
interesting decay channels to establish the NMSSM effects which are
absent in the MSSM. 
In this context, searching for lighter Higgs bosons, 
in particular $A_1$ is very interesting, since it can be 
very light~\cite{Mahmoudi:2010xp,Bomark:2014gya}. 
There are many phenomenological analysis
reported in the literature exploring the detection prospect of 
$A_1$ at the 
LHC~\cite{Belyaev:2008gj,Belyaev:2010ka,Almarashi:2011hj,Cerdeno:2013cz,
Curtin:2014pda,Bomark:2015fga}.
In our study as
reported in \cite{Guchait:2015owa}, the rates of production of
non SM-like Higgs bosons in various decay channels are estimated for the
LHC Run 2 experiment with the center of mass energy, $\sqrt{S}=$13~TeV.
Remarkably, it is observed that along with the dominant $b \bar b$ and
$\tau\tau$ decay modes of non SM-like Higgs bosons, the 
BR for two photon $(\gamma\gamma)$ decay mode is also 
very large for a certain part of the parameter space. 
In particular, light $A_1$ decays to $\gamma\gamma$ mode
with a BR ranging from a few percent
to 80-90\% for a substantial region of the parameter
space\cite{Arhrib:2006sx,Dermisek:2007yt,Kim:2012az,Christensen:2013dra,Ellwanger:2015uaz,Guchait:2015owa,Bomark:2015fga}. 
On the other side, as we know, experimentally photon is a very clean 
object and can be reconstructed with a very high precision, which motivates 
us to study the signal of non SM-like Higgs boson 
in this $\gamma\gamma$ 
channel~\cite{Moretti:2006sv,Bomark:2014gya,Bomark:2015fga}.
In this context, it is to be noted that, 
neither the SM nor the MSSM predict
this large rate of $\gamma\gamma$ decay mode of any of the Higgs boson 
for any region of the parameter space.
Hence, this distinct feature appears to 
be the characteristic signal of the NMSSM and can be exploited 
in distinguishing it from
the other SUSY models. More precisely, in the presence of any SUSY signal, 
this di-photon decay mode of $A_1$ can be used as a powerful avenue 
to establish the type of the SUSY model.

%%%%%%%%%%%%%%%%%%%%%%%%%%%%%%%%%%%%%%%%%%%%%%%
In this present study, mainly we focus on $A_1$ and 
explore its detection possibility in the $\gamma\gamma$ mode.
In principle, $A_1$ can be produced directly via the standard SUSY 
Higgs production mechanisms, i.e primarily via the gluon gluon fusion 
or through $b$ and $\bar b$ annihilation. 
However, in both the cases, the production cross sections are
suppressed due its singlet nature.
In our study, we employ the SUSY particle 
production,  namely the associated chargino-neutralino  
and, the subsequent decay of heavier neutralino state produces $A_1$, 
followed by $A_1 \to \gamma\gamma$ decay.
The combination of lighter chargino($\tilde\chi_1^\pm$) and, 
either of the second ($\N0_2$) or the third ($\N0_3$) 
neutralino states is found to be produced dominantly at the LHC 
energy~\cite{Ghosh:2012mc,Cerdeno:2013qta}.
In the final state, in order   
to control the SM backgrounds, we require also one associated lepton 
arising from $\C1$ decay. The production and decay mechanism of the entire
process is shown as, 
\br
 pp \rightarrow \decay{\C1}{\N0_1 \ell^\pm\nu} ~~~ + ~~~~
\decay{\N0_{j}}{\N0_1 ~\decay{A_1}{\gamma\gamma}},  \ \ (j=2,3)
\label{eq:1}
\er
\begin{center}
\begin{figure}[t]
  \includegraphics[height=5.0cm,width=18.0cm,trim={-2cm 21cm 0 3.8cm},clip]{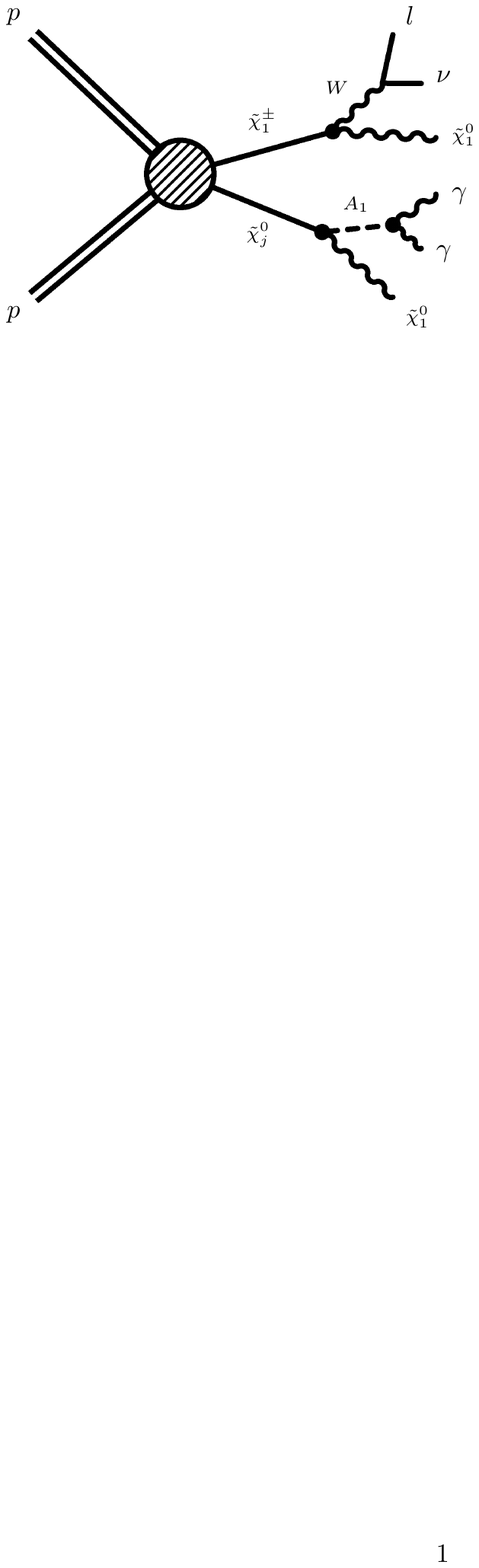}
\caption{\small Lighter chargino($\C1$)-neutralino($\N0_j$),
(j=2,3) associated production in proton-proton collision followed by 
cascade decays to two photons and a lepton along with 
lightest neutralinos, as Eq.~\ref{eq:1}.}
\label{fig:cascade}
\end{figure}
\end{center}

\vspace{10cm}
schematically it is presented in Fig.~\ref{fig:cascade}. 
The final state contains
hard missing energy due to the presence of neutrinos and  
neutralinos($\N0_1$) which are assumed to be the lightest SUSY particle(LSP) 
and stable \footnote{We are considering R-Parity conserving model.}, and 
escape the detector, since they are weakly interacting.
Finally, the reaction,Eq.~\ref{eq:1} leads to the signal, 
\br
 {\gamma\gamma + \ell^{\pm} + \slashed E_T}. 
\label{eq:sig}
\er
Of course, in addition to the chargino-neutralino  production cross section, 
the BR($\N0_{2,3} \to \N0_1 A_1$) and BR($A_1 \to \gamma\gamma$), which 
are sensitive to the parameter space, very crucial in determining 
the signal rate. In view of this, we investigate the sensitivity of 
this signal to the relevant parameters scanning those systematically 
for a wide range and  
identify the suitable region which provides the reasonable rate of the  
signal. Finally, out of this parameter scan, 
we select few benchmark parameter points for which 
results are presented.  
Performing a detail simulation including detector effects for  
both the signal and the SM backgrounds processes, we predict 
the signal significances corresponding to our choices of 
parameters for a few integrated luminosity options at the LHC with the 
center of mass energy, $\sqrt{S}$=13~TeV.

This paper is organized as follows, In section 2,  
after briefly discussing the chargino and neutralino sector in the NMSSM, 
we study the parameter space sensitivity of  
chargino-neutralino associated production cross section.
The parameter sensitivity of BRs of neutralinos and $A_1$ decays 
are discussed in section 3 and then propose
few benchmark points for which results are presented. 
The details of the simulation are 
presented in section 4, while results are discussed in section 5.
Finally, we summarize in section 6.
%%%%%%%%%%%%%%%%%%%%%%%%%%
\section{Chargino-Neutralino production}
The chargino-neutralino associated 
production($\tilde\chi^\pm_1\tilde\chi_{2,3}^0$) 
in proton-proton collision is mediated purely by electro-weak(EW) 
interaction at the tree level and, hence very sensitive to the parameters 
space owing to the 
dependence of couplings.   
Therefore, in order to understand the various features of this production    
process at the LHC, it is worth 
to discuss the interplay between parameters and cross sections.

\subsection{Chargino and Neutralino sector in NMSSM}
In SUSY model, there are spin half EW gauginos 
and Higgsinos which are the supersymmetric partners of the 
gauge bosons and Higgs bosons respectively. 
The soft mass terms for gauginos and the spontaneous breaking of EW
symmetry lead a mixing between gaugino and Higgsino states making 
them weak eigenstates without physical mass terms.
The charginos are the mass eigenstates corresponding to the 
mixed charged gaugino and Higgsino states.  
Similarly, the mixings of 
neutral EW gauginos and Higgsinos produce 
physical neutralinos.    
The masses and the corresponding physical states 
can be obtained by diagonalizing the respective mass matrices.
For instance, the masses of the chargino states ($\tilde \chi_{1,2}^{\pm}$) 
are obtained diagonalizing the $2 \times 2$ chargino mass matrix 
by a bi-unitary transformation. 
In the MSSM, the masses and composition of these chargino states 
are determined by $M_2$ - the $SU(2)$ gaugino mass parameter,
$\mu$ and $\tan\beta$ - the ratio of two vacuum expectation 
values($v_u,v_d$) of 
the neutral components of two Higgs doublets require to break EW symmetry 
spontaneously. In the NMSSM, the presence of an extra 
Higgs singlet field does not modify the chargino sector, 
hence it remains same as in the MSSM, except the Higgsino mass parameter 
$\mu$ which is replaced by $\rm \mu_{eff}$. 

On contrary, in the NMSSM, the neutralino sector 
is extended due to the addition of an extra singlino state $\tilde S$ - 
the fermionic superpartner of the singlet scalar field $(S)$.
Here $\tilde S$ mixes with the Higgsinos due to the presence of the
$\lambda H_u H_d S$ term in the superpotential. Thus, the resulting 
${\bf 5 \times 5}$ neutralino mass matrix is given by,
\begin{equation}
\rm M_N = \left( \begin{array}{ccccc}
M_1 &0 &\frac{-g_1 v c_{\beta}}{\sqrt 2} & \frac{g_1 v s_{\beta}}{\sqrt 2}  & 0  \\
0 &M_2   & \frac{g_2 v c_{\beta}}{\sqrt 2} & \frac{-g_2 v s_{\beta}}{\sqrt 2} &0 \\
\frac{-g_1 v c_{\beta}}{\sqrt 2} & \frac{g_2 v c_{\beta}}{\sqrt 2}  & 0 & - \mu_{eff} &  -\lambda v s_{\beta}\\
\frac{g_1 v s_{\beta}}{\sqrt 2}   & \frac{-g_2 v s_{\beta}}{\sqrt 2}  & - \mu_{eff}  & 0 &  - \lambda v c_{\beta} \\
0 & 0 &  -\lambda v s_{\beta} &  - \lambda v c_{\beta}  &  2 \kappa v_s \end{array} \right).  
\end{equation}
Here $M_1$ is the mass of $U(1)$ gaugino -  
the bino($\tilde B$) and $g_1$, $g_2$ are the
weak gauge couplings. 
In the MSSM limit, \emph{i.e.} $\lambda,\kappa \rightarrow 0$,
this $5 \times 5$ neutralino mass matrix reduces to a $4 \times 4$ 
mass matrix. 
The masses of neutralinos can be derived by diagonalizing symmetric matrix 
$\rm M_N$ via a unitary 
transformation as,
\br
 \rm M_{\tilde \chi^0}^D  = \rm N M_N N^{\dagger}.
\label{eq:nn}
\er 
with N as a unitary matrix.
The analytical solution of the neutralino mass matrix presenting the spectrum
of neutralino masses and mixings
exist in the literature for the MSSM~\cite{Guchait:1991ia,Choi:2001ww}. 
However for the NMSSM, the 5th order eigenvalue equation makes it more 
difficult to extract exact analytical solution. Nevertheless, attempts 
are there to find the approximate analytical 
solution~\cite{Pandita:1994vw,Choi:2004zx}.
Consequently, the five physical neutralino states become the admixtures of
weak states, such as gauginos, Higgsinos and singlino. 
Hence, in the basis $\tilde \psi^0 \equiv (-i\tilde{B},-i\tilde{W_3},
\tilde{H_d^0}, \tilde{H_u^0},\tilde{S})$, the physical neutralino 
states are composed of,
\br
 \rm \tilde \chi_i^0  = \rm N_{i j} \tilde \psi_j^0,
\label{eq:nmix}
\er
where $\rm N_{ij}$(i,j=1-5) is defined by Eq.~\ref{eq:nn}.
In particular, $\rm N_{i5}$ presents the singlino component  
in the $i$-th physical neutralino state.
 To conclude, in the NMSSM, the masses and the mixings of the 
charginos and neutralinos at the tree level can be determined by 6
parameters, namely,
\br
M_1, \ \ M_2, \ \ \tan \beta, \ \ \mu_{eff}, \ \ \lambda, \ \ \kappa.
\er
Here one can choose $M_1$ and $M_2$ to be real and
positive by absorbing phases in $\tilde B^0$ and $\tilde W^0$ respectively,
but in general $\mu_{eff}$ can be complex. In this current study,
we assume CP-conserving NMSSM setting all the input parameters real.

A careful examination of the neutralino mass matrix reveals
few characteristic features of this sector~\cite{Pandita:1994vw,Choi:2004zx}.
For instance, notice that the singlet field does not mix with the
gauge fields, and hence the singlino like neutralino states do not
interact with the gaugino like states or gauge fields. 
Apparently, two out of the five neutralino states remain to be
gaugino like if, $|M_{1,2}-\mu_{eff}|> M_Z$. Note that the direct
singlet-doublet mixing is determined by $\lambda$. 
The mass of the singlino like neutralino is given by $|2 \kappa v_s|$, 
and so if $|2\kappa v_s| << M_{1,2},\mu_{eff}$, then the lighter 
neutralino state becomes dominantly a singlino like.
On the other hand, if $|2 \kappa v_s| >> M_{1,2}, \mu_{eff}$, then the 
singlino state completely decouples from the other states resulting all 
four neutralino states mixtures of gaugino-Higgsino, \emph{i.e} a 
MSSM like scenario, where as the remaining heavier neutralino state 
appears to be completely singlino like.
The coupling structures of neutralinos
with gauge bosons and fermions remain the same as in the MSSM,
since the singlet field does not interact with them. 
For the sake of discussion in the later section, 
we present the $\rm {\C1-\N0_j-W^\mp}$ interaction,
\begin{equation}
g_{\tilde\chi_1^{\pm}\tilde \chi_j^0 W^{\mp}}^{L}
	= \frac{ e}{s_w} \left (N_{j2} V_{11}^* - \frac{1}{\sqrt 2} N_{j4} 
V_{12}^*  \right), ~ g_{\tilde\chi_1^{\pm}\tilde \chi_j^0 W^{\mp}}^{R}
	= \frac{ e}{s_w} \left (N_{j2}^* U_{11} + \frac{1}{\sqrt 2} N_{j3}^* 
U_{12}  \right),
\label{eq:cnw}
\end{equation}
and $\rm {q - \tilde q -\N0_j}$ couplings,
\begin{equation}
g_{d \tilde d \chi_j^0 }^{L} 	\approx
\frac{-e}{\sqrt{2} s_w c_w} \left ( \frac{1}{3} N_{j1} s_w -N_{j2} c_w  \right ),
\ \   g_{ d \tilde d \tilde \chi_j^0 }^{R} 	\approx 0,
\label{eq:ddx}  
\end{equation}
\begin{equation}
g_{u \tilde u \tilde \chi_j^0 }^{L} 	\approx
\frac{-e}{\sqrt{2} s_w c_w} \left ( \frac{1}{3} N_{j1} s_w +N_{j2} c_w  \right ),
\ \   g_{u \tilde u \tilde \chi_i^0 }^{R} 	\approx 0.
 \label{eq:uux} 
\end{equation}
with $\rm s_w= \rm \sin\theta_w$,$\rm c_w=\rm \cos\theta_w$ and $j$=2,3
Note that, since we consider only the first two generations of squarks 
and assume that the chiral mixings are negligible, hence we 
omit the corresponding interaction terms and, for the same reasons, 
$g_{u \tilde u \tilde \chi_j^0 }^{R}$ and 
$g_{d \tilde d \tilde \chi_j^0 }^{R}$ 
are negligible. Apparently, 
the presence of the direct effect of NMSSM through singlino component is 
absent in these interactions. However, because of the unitarity of 
the mixing matrix $\rm N$, the singlino component $N_{i5}$ indirectly 
affects these couplings. It will be discussed more in the next 
sub-section in the context of the chargino-neutralino production.

\subsection{$\C1\N0_j$ cross-section }
In this section, in the framework of the NMSSM, we discuss various
features of the chargino-neutralino($\C1\N0_j$,j=1,2,3) 
associated production at the LHC. For the sake of comparison and 
discussion, we also study $\C1\N0_1$ production cross section, although
it has no relevance to our present context. 
\begin{figure}[t]
\centering
  \includegraphics[height=7.0cm,width=15.0cm,trim={0cm 18cm 0 3.8cm},clip]{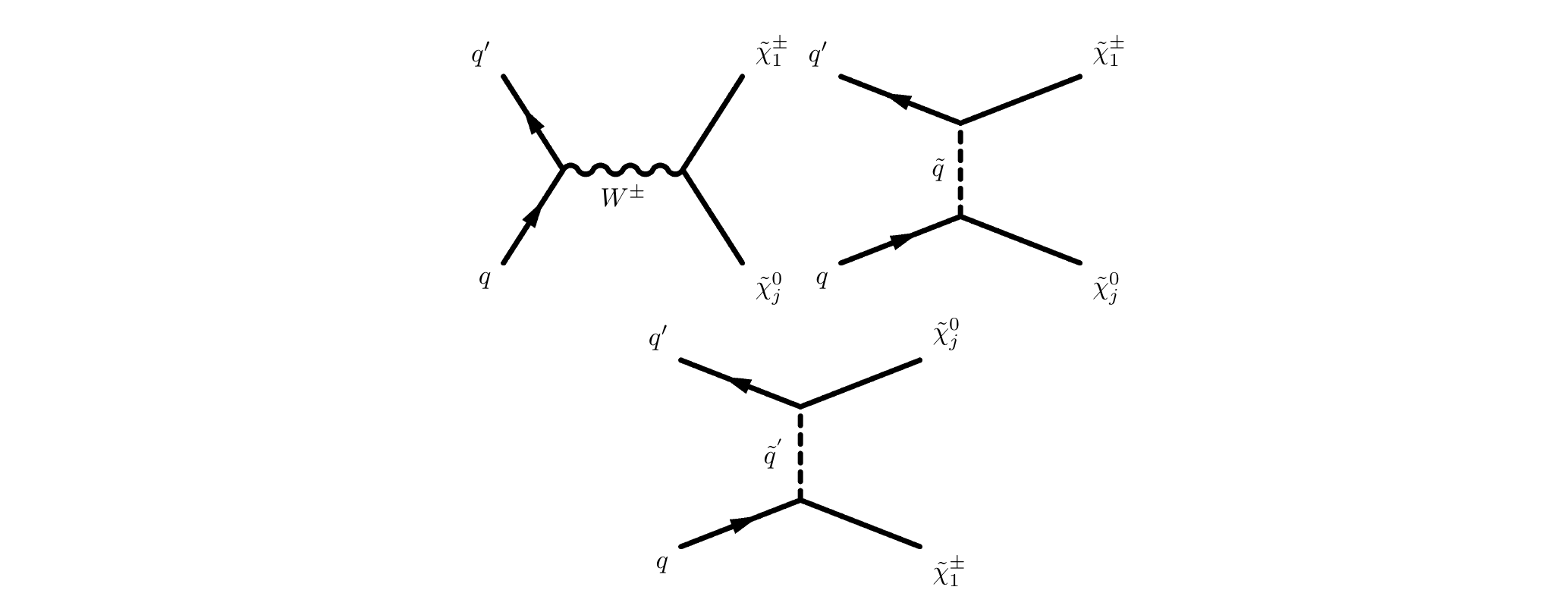}
\caption{\small Tree level Feynman diagrams for chargino-neutralino 
associated production via q and $\rm \bar q'$ annihilation.}
\label{fig:x2x1}
\end{figure}
As already mentioned, in hadron colliders, the chargino-neutralino pairs 
are produced purely via EW interaction initiated by quark 
and anti-quark annihilation as,
\br
 q \bar q' \rightarrow \tilde\chi_1^\pm \N0_j; \ \ \  j=1,2,3, 
\er
the corresponding Feynman diagrams at the tree level are shown 
in Fig.~\ref{fig:x2x1}.
The $s$ and $t/u$-channels are mediated 
by the $W$ boson and the first two generations of squarks respectively
and, are very sensitive to the couplings, see Eq.~\ref{eq:cnw}--\ref{eq:uux}, 
which are regulated by model parameters. 
In case, if both the chargino and the neutralino states be pure 
Higgsino like, 
then the $t$ and $u$ channel diagrams decouple completely due to the 
suppressed quark-squark-neutralino
couplings(Eqs.~\ref{eq:ddx},~\ref{eq:uux}), otherwise 
mixed or pure gaugino likes states are favored.
The contribution of the $t/u$-channel diagrams are also suppressed for
heavier masses of squarks. Moreover, negative interference of the  
$s$ and $t/u$ channel diagrams yields 
an enhancement of the production cross section for heavier
masses of squarks for a given set of other parameters.

The partonic level differential $\C1\N0_j$ cross section in NMSSM 
can be obtained following the form given 
in  Ref.~\cite{Beenakker:1999xh} for the MSSM,
\begin{align}
\frac{d \hat \sigma (q \bar q \rightarrow \tilde \chi_i^{\pm} 
\tilde \chi_j^{0} )}{d\hat t} ~&= ~
\frac{\pi \alpha^2}{3 \hat s^2} 
\big [~|Q_{LL}|^2 ~(\hat u -m_{\tilde\chi_{j}^0}^2) ~(\hat u -m_{\tilde \chi_{i}^-}^2)~ 
+~|Q_{LR}|^2 (\hat t -m_{\tilde\chi_{j}^0}^2)(\hat t -m_{\tilde\chi_{i}^-}^2)~ \nonumber \\
& ~+~ 2~\hat s ~Re(Q_{LL}^* Q_{LR}) ~m_{\tilde \chi_j^0} ~m_{\tilde \chi_i^-}  \big ] 
\end{align}
which is expressed in terms of four helicity charges 
$ Q_{LL}, Q_{LR}, Q_{RL}, Q_{RR}$. 
For the sake of completeness, we also present the explicit form of 
these charges \cite{Beenakker:1999xh},
\br
Q_{LL} &=& \frac{1}{\sqrt{2}s_w^2} \left [ \frac{N_{j2}^{*} V_{i1}
- 1/\sqrt{2} ~N_{j4}^* V_{i2}}{\hat s-M_W^2 }
~ + ~ V_{i1} \frac{ I_{3 \tilde q } N_{j2}^* +(e_{\tilde q} -I_{3 \tilde q}) N_{j1}^* \tan \theta_w }{\hat u - m_{\tilde q}^2} \right ],   \nonumber
\\
Q_{LR} &=& \frac{1}{\sqrt{2}s_w^2} \left [ \frac{N_{j2}
U_{i1}^* + 1/\sqrt{2} ~N_{j3} U_{i2}^*}{\hat s-M_W^2 }
~ - ~(U_{i1})^* \frac{ I_{3 \tilde q^{'} } N_{j2} + (e_{\tilde q^{'}} -
I_{3 \tilde q^{'}}) N_{j1} \tan \theta_w}{\hat t - m_{\tilde q^{'}}^2}         \right ],
\nonumber
\\
Q_{RR}&=&Q_{RL} =0,
\er
where the Mandelstam variables are defined as, 
$\hat s = (p_1 +p_2)^2; \ \ \hat t = (p_1-p_3)^2;\ \ \hat u = (p_2-p_4)^2$
in the partonic frame,
$p_1, p_2$ are the momenta of initial quarks, $p_3,p_4$ represent the
same for $\tilde\chi^\pm_i$ and $\N0_j$ respectively.
Notice that, as pointed out earlier, even without any explicit
dependence of couplings, Eqs.~\ref{eq:cnw},\ref{eq:ddx} and \ref{eq:uux}, 
on the singlino composition, $\rm N_{j5}$ in the 
neutralino state, nonetheless, it affects the
$\C1\N0_j$ production cross section due to the dilution of  gaugino and
Higgsino components.

We compute this leading order(LO) cross section setting 
QCD scales, $Q^2 = \hat s$-the partonic center of mass energy 
and for the choice of CT10~\cite{Lai:2010vv} parton distribution function. 
The corresponding next to leading order(NLO) predictions for the 
$\C1\N0_j$ cross sections 
are obtained from Prospino \cite{Beenakker:1996ed} and 
the k-factor(=$\rm \sigma_{NLO/\sigma_{LO}}$ is found to 
be $\sim 1.3$ \cite{Beenakker:1999xh}. 
In the present NMSSM case, to take care NLO effects in the 
cross-section, we use the same k-factor, which is not expected to be too 
different with respect to the MSSM case.   
We observe that LO chargino-neutralino associated production 
cross-section varies from sub femto-barn(fb) level to  
to few pico-barn(pb) for the mass range of 100-500 GeV of 
charginos and neutralinos.  

\begin{figure}[t]
\hspace{0.2cm}
\includegraphics[height=7.3cm,width=7.3cm]{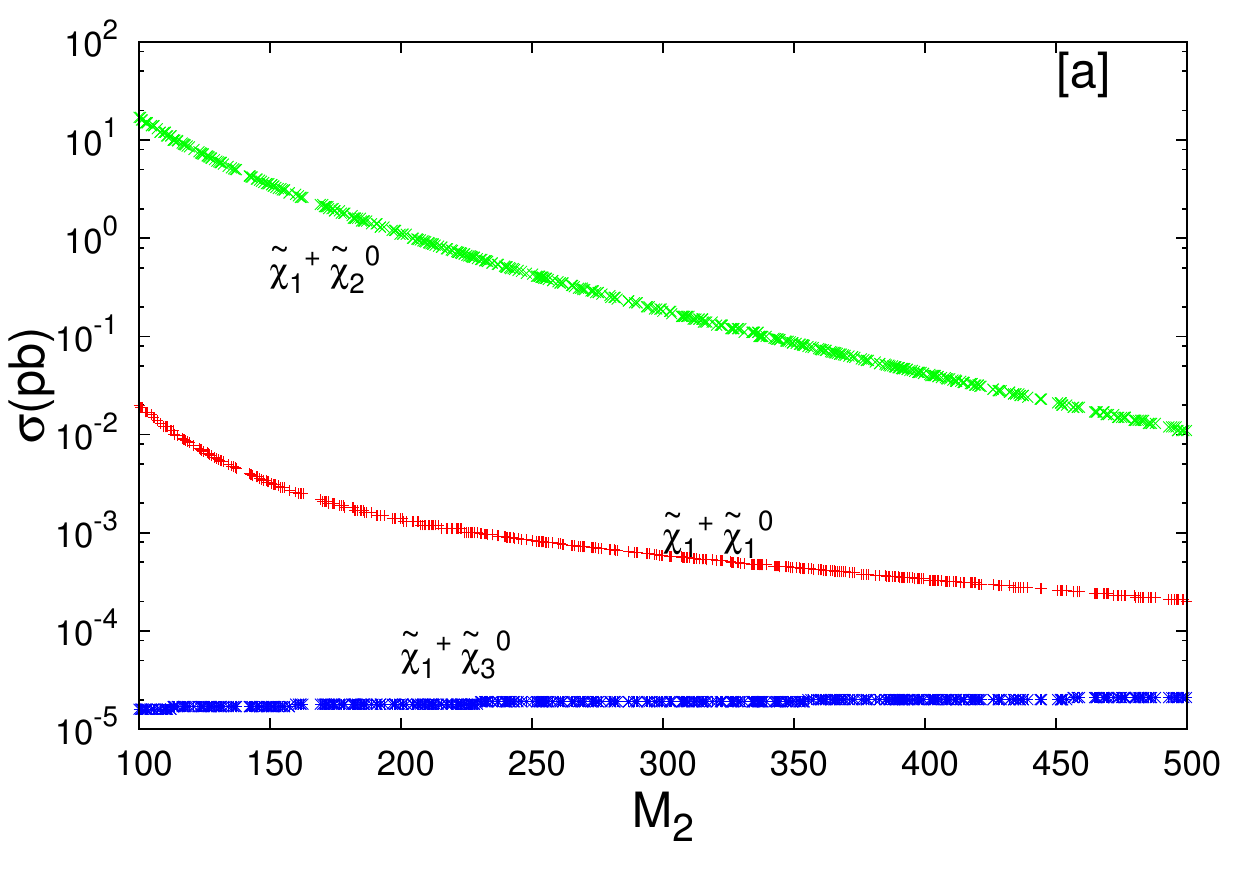}
\hspace{-0.1cm}
\vspace{-0.4cm}
\includegraphics[height=7.3cm,width=7.3cm]{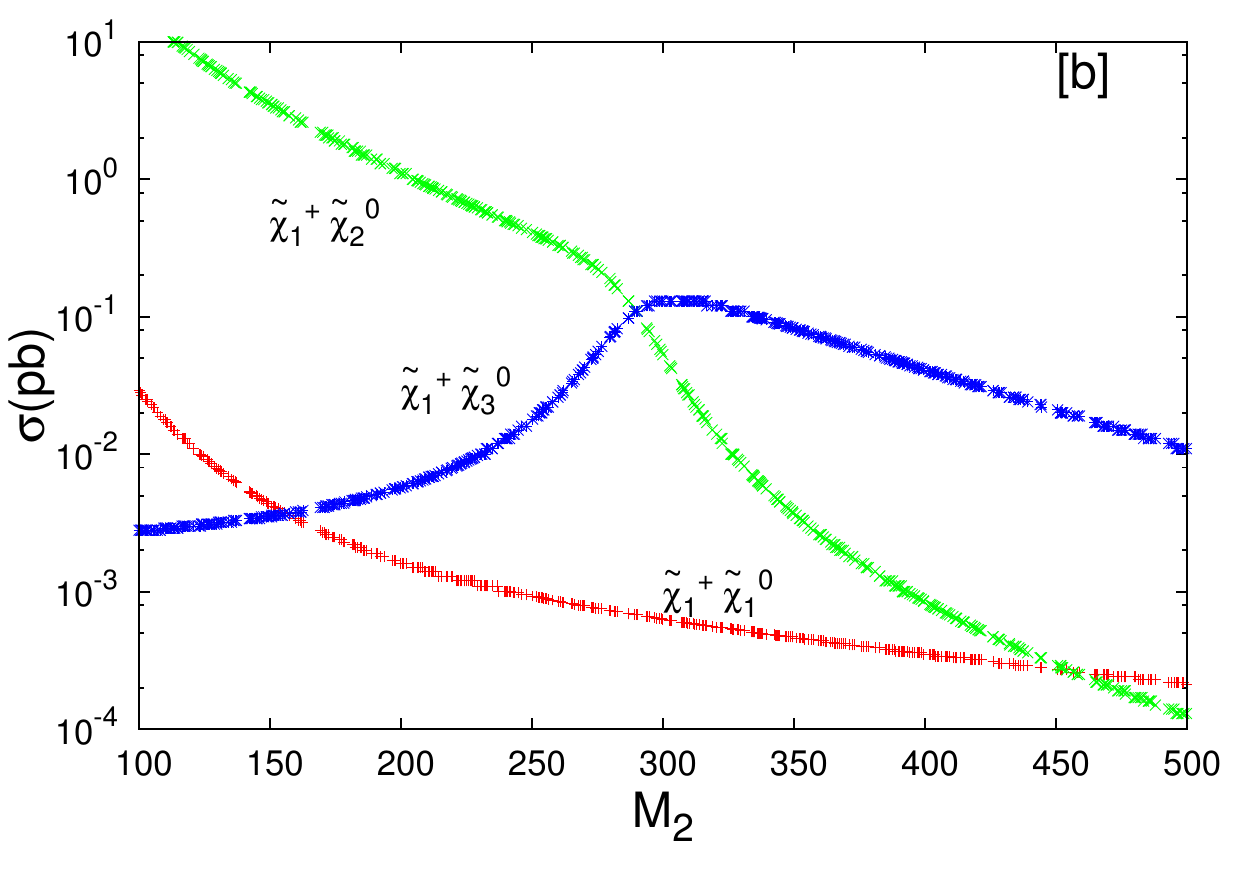}
\caption{\small Variation of leading order (LO) chargino-neutralino 
associated production
cross section with $M_2$, at the LHC energy $\sqrt{S}$ = 13~TeV and 
for two choices
of $\lambda, \kappa$=[a]~0.1,0.7, [b]~0.7,0.1.
The other parameters
are set as, $\rm \mu_{eff}=1000~GeV, \rm M_1=M_2/2$, $\tan\beta= 10$.}
\label{fig:csm2}
\end{figure}

\begin{figure}[t]
\includegraphics[height=7.0cm,width=7.0cm]{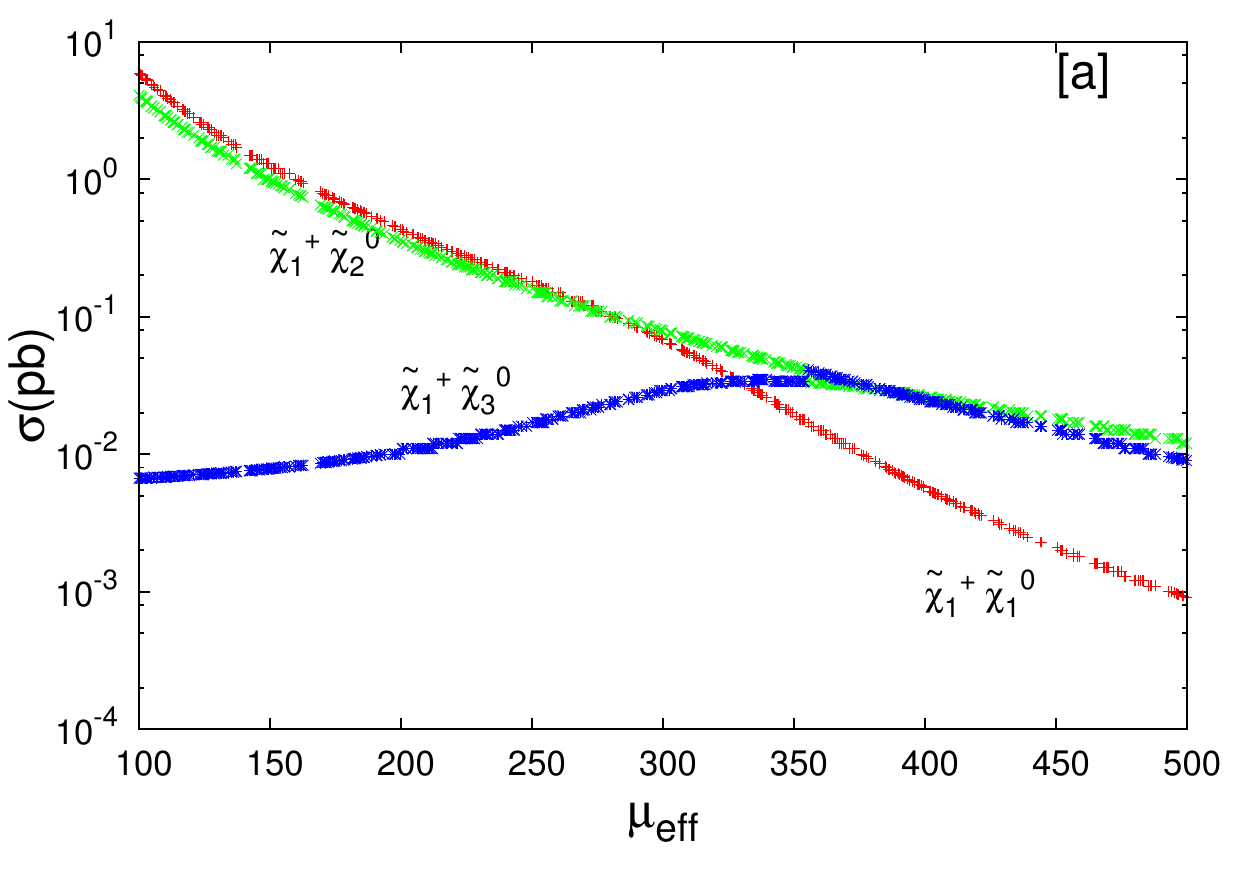}
\hspace{-0.1cm}
\vspace{-0.45cm}
\includegraphics[height=7.0cm,width=7.0cm]{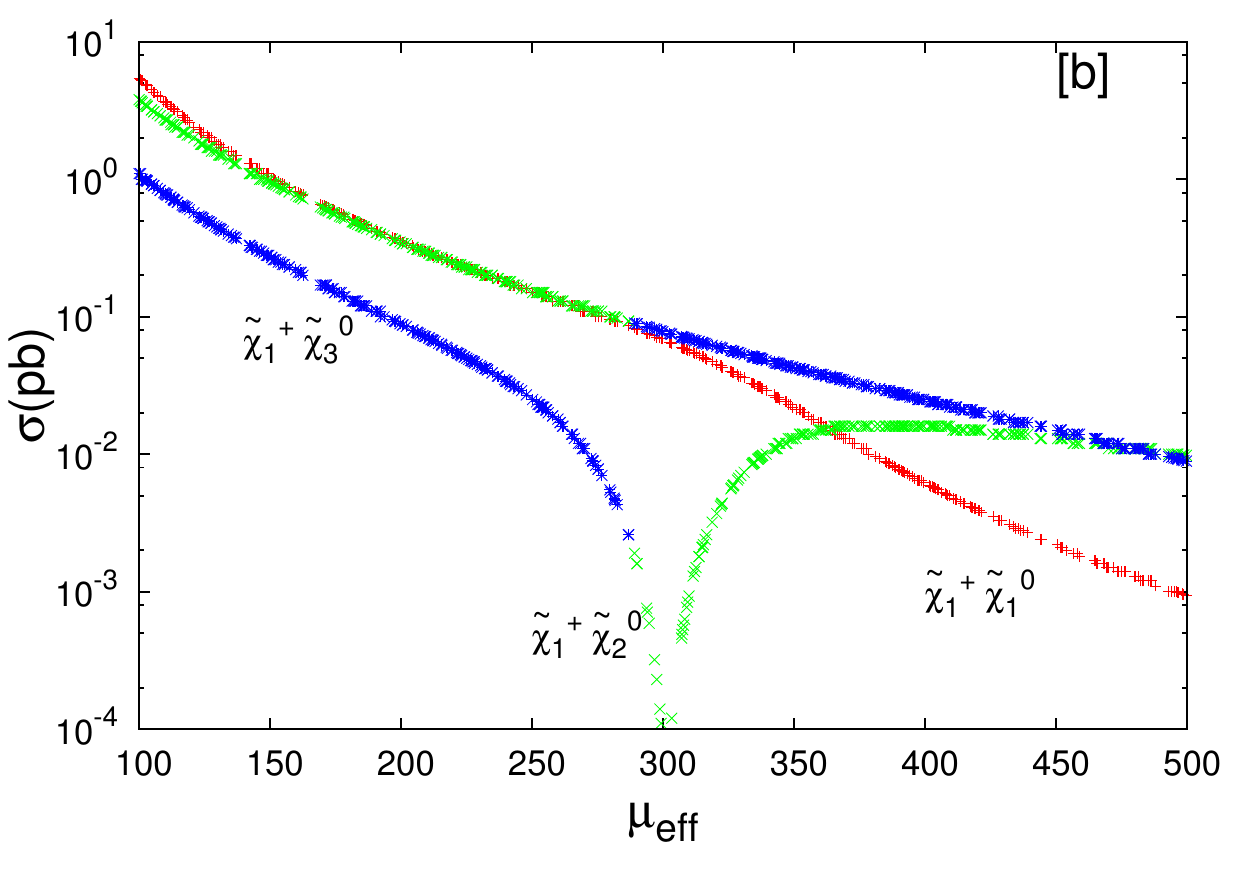}
\includegraphics[height=7.0cm,width=7.0cm]{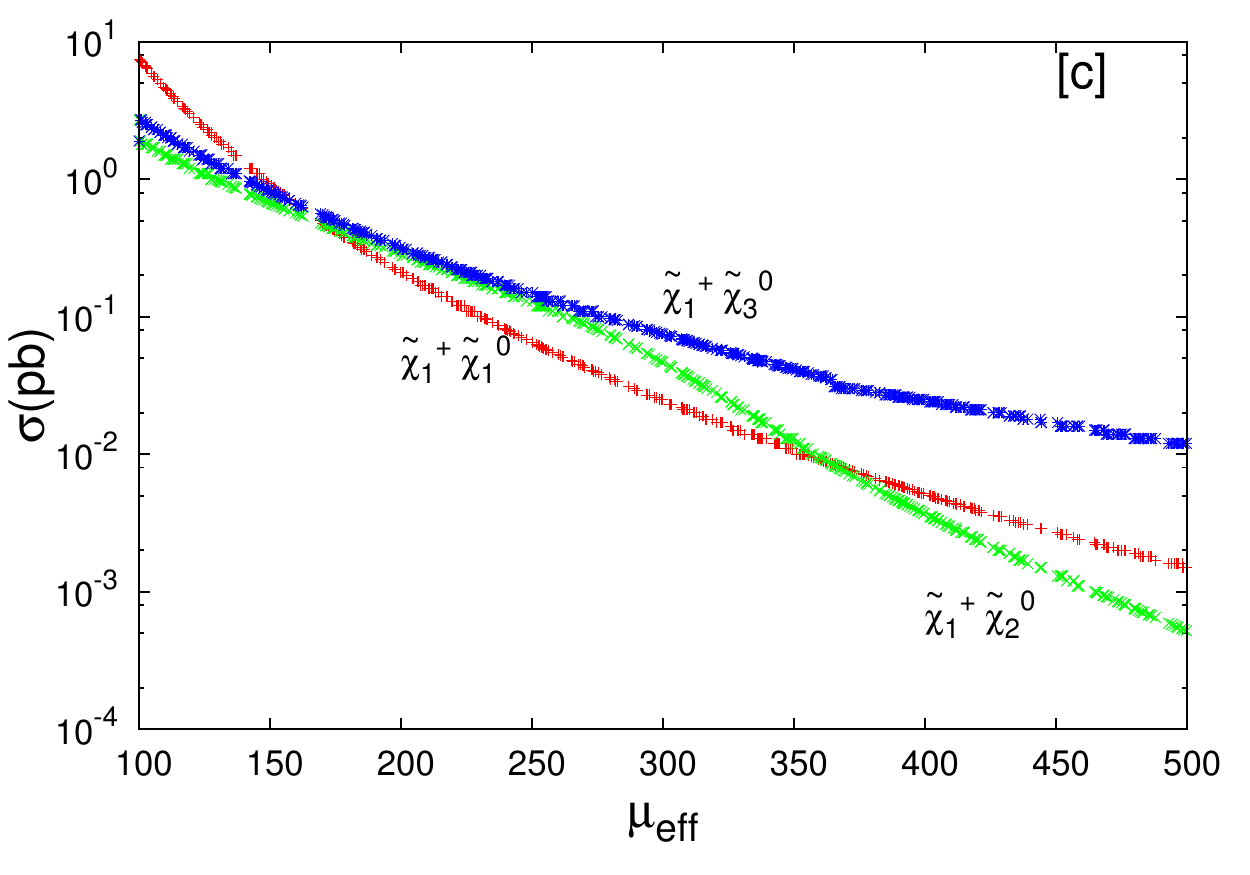}
\centering
\caption{\small Variation of LO chargino-neutralino associated production
cross section with $\mu_{eff}$, at the LHC energy $\sqrt{S}$=13~TeV
and for the choices of $\lambda,\kappa$=[a]~0.1,0.7, [b]~0.2,0.1,[c]~0.4,0.1.
The other parameters are set as, $\rm M_2$=600~GeV, $\rm M_1= \rm M_2/2$,
$\tan \beta= 10$.
}
\label{fig:csmu}
\end{figure}

To understand the dependence of $\tilde\chi_1^\pm \tilde\chi_{j}^0$ 
cross sections on the parameters,
we demonstrate its variation in Fig.~\ref{fig:csm2} and Fig.~\ref{fig:csmu},
primarily for gaugino and
Higgsino like scenarios 
varying $\rm M_2$ and $\rm \mu_{eff}$ respectively.
The variation of singlino composition are controlled by a set of few 
choices of $\lambda, \kappa$= [a]~0.1,0.7, 
[b]~0.2,0.1 for Fig.~\ref{fig:csm2} and 
$\lambda,\kappa=$[a]~0.7, 0.1, [b]~0.2, 0.1 and [c]~0.4, 0.1 for 
Fig.~\ref{fig:csmu}.
The other parameters are set as, $\tan\beta=10$, 
$\rm\mu_{eff}=$1000~GeV(for Fig.\ref{fig:csm2}), 
$\rm M_2$=600~GeV(for Fig.~\ref{fig:csmu}),
squark masses $\rm m_{Q_{L}}, m_{D_{L,R}}=1000$~GeV and 
assuming the relation $\rm M_1 = M_2/2$. 
In the following, we discuss the variation of cross sections 
with the sensitive parameters
which has some impact on the signal sensitivity, as
will be discussed in the later sections.

{$\bul$} The dependence of $\C1\N0_j$ cross section on $M_2$, in the gaugino 
like scenario($M_2 < \mu_{eff}= 1000~GeV$) is 
presented in Fig.\ref{fig:csm2}. In this scenario, 
in the case of $\lambda, \kappa=[a]~0.1,0.7$, the mass of singlino 
is very heavy
($\rm \sim |2 \kappa v_s| = \rm 2\mu_{eff} \kappa/\lambda= \rm 14~TeV$)
and  the $\C1$ state is wino like of mass around $\rm M_2$,
while the $\N0_1$ is bino dominated with 
its mass about $m_{\N0_1}\sim M_1$. 
On the other hand, because of large mass of the singlino state and 
lower value of $\lambda$, \emph{i.e} small singlet-doublet mixing, 
the $\N0_2$ and $\N0_3$ states are turn out to be 
dominantly wino and Higgsino like
respectively, with masses $\rm m_{\N0_2} \sim M_2$ and
$\rm m_{\N0_3} \sim \mu_{eff}$.
It explains the reasons of larger $\C1\N0_2$ 
cross section in comparison to $\C1\N0_1$, as seen in Fig.~\ref{fig:csm2}[a]. 
Note that, the subsequent fall of both the cross sections 
with the increase of $M_2$ is purely a mass effect.
Obviously, the $\C1\N0_3$ cross section is expected to be suppressed
and almost negligible dependence on $\rm M_2$. 
However, in the case of $\lambda, \kappa$=[b]~0.7,0.1, 
the singlino state becomes comparatively light with mass about
$\sim $300~GeV. 
In this scenario, due to the large singlet-doublet mixing $(\lambda =0.7)$, 
at the lower values of $M_2$, the $\N0_3$ state is found to be singlino like 
with very less wino and Higgsino components, whereas $\N0_2, \N0_1$ 
states appear to be more or less wino and bino like respectively. 
Consequently, in this lower region of $M_2$, the   
$\C1\N0_{1,2}$ cross sections are   
higher than the $\C1\N0_3$, mainly due to the
suppressed couplings of $\N0_3$ with 
gauge boson and fermions being it a dominantly a singlino state. 
However, with the increase of $M_2$, the wino 
(singlino) component in $\N0_2$ ($\N0_3$) decreases, 
resulting a gradual fall(enhancement) of $\C1 \N0_2$ 
($\C1 \N0_3$) cross sections.
Eventually, as $M_2$ reaches closer to $|2 \kappa v_s| \sim 300$~GeV, 
the $\N0_2$ and $\N0_3$ states
tend to be singlino and wino like respectively and, hence due to   
the depletion of $ \C1\N0_2 $ cross section very sharply,  
$\C1\N0_3$ cross section takes over it and then  
falls slowly mainly due to the phase space suppression, see
Fig.~\ref{fig:csm2}[b].
However, in contrast, due to the larger mass of 
singlino($\sim$ 14 TeV) the similar type of crossing between 
$\C1\N0_2$ and $\C1\N0_3$ cross sections is not observed 
in Fig.~\ref{fig:csm2}[a].

{$\bullet$} The variation of cross sections with $\mu_{eff}$, 
for Higgsino like scenario 
is presented in Fig.\ref{fig:csmu}, 
keeping $\rm M_2=$600 GeV and for three combinations  
of $\lambda$, $\kappa$= [a]~0.1,0.7, [b]~0.2, 0.1, [c]~0.4, 0.1. 
In this scenario, the $\C1$ state is mostly Higgsino like for the lower range
of $\rm \mu_{eff}$, and then becomes a gaugino-Higgsino mixed 
state when $\mu_{eff} \sim M_2$.
For the scenario [a], at the 
lower range of $\mu_{eff}$($\lesssim M_1$=300~GeV), 
the Higgsino composition in $\N0_1$ state  
is the dominant one, but it becomes 
bino like once $\rm \mu_{eff} \gtrsim M_1$ and a     
drop of $\C1\N0_1$ cross section occurs beyond $\rm \mu_{eff} \sim$300~GeV,
as seen in Fig.~\ref{fig:csmu}[a]. 
However, for the scenario, [b] and [c], at the lower side of 
$\rm \mu_{eff}$, the $\N0_1$ state, along with some Higgsino component, 
contains a finite fraction of singlino 
(recall the singlino mass $\sim \rm 2\mu_{eff}\kappa/\lambda$), and 
in particular, for the scenario[c], $\N0_1$ becomes dominantly a singlino like.
Nevertheless, the $\C1\N0_1$ cross section are not heavily suppressed 
due to the presence of mild Higgsino component in the $\N0_1$ state.
The Higgsino and bino like nature of $\N0_2$ yields a steady 
variation of $\C1\N0_2$ cross section with $\mu_{eff}$, except for the
case [b] where a sudden drop and then further an enhancement
is observed at $\mu_{eff} \sim$ 300 GeV.
Here both the singlino and the bino masses are around $\sim$300 GeV, 
implying an increase of singlino and bino components in 
$\N0_2$ state causing a drop
of $\C1\N0_2$ cross section and beyond  
this region, again it goes up with the increase of 
$\rm \mu_{eff}$ due to further increase of its Higgsino component.     
In the presence of small singlet-doublet mixings, in the scenario 
$\lambda, \kappa$ =[a]~0.1,0.7, the $\N0_3$ state is bino dominated 
at the lower range of  
$\rm \mu_{eff}< M_1$, resulting comparatively a lower $\C1\N0_3$ 
cross section, which
slowly increases with $\mu_{eff}$ due to the enhancement of 
Higgsino composition in it, 
as observed in Fig.\ref{fig:csmu}[a].  In  Fig.\ref{fig:csmu}[b], 
it is found that the singlino composition
in $\N0_3$ state goes up with the increase of $\rm \mu_{eff}$,
while it is below $|2 \kappa v_s|$
and, becomes completely singlino like at 
$\rm \mu_{eff}\sim |2 \kappa v_s|$ $(\sim 300 GeV)$ 
hence the rapid fall of $\C1 \N0_3$ cross section.
Beyond $\rm\mu_{eff}>2|\kappa v_s|$ region, Higgsino composition in the 
$\N0_3$ state increases yielding more higher $\C1\N0_3$ cross
section and then due to mass effect, it falls slowly.  
%%%%%%%%%%%%%%%%%%%%%%%%%%%%%%%%%%%%%%%%%%%%%%%%%%%%%%5
\section
{Decays : $\tilde \chi_{2,3}^0 \to \N0_1 A_1; A_1 \rightarrow \gamma \gamma$}

As stated earlier, the sensitivity of the signal
$\ell +\gamma\gamma + \PMET$, crucially depends on the combined effects of 
the $\tilde \chi^{\pm}_1 \N0_{2,3}$ production cross
section and subsequent BRs involved in the 
cascade decays, such as $\N0_{2,3} \to \N0_1 A_1$ 
and $A_1 \to \gamma\gamma$,  
$\C1 \to \N0_1 \ell\nu$. 
Note that the BR($\C1 \to \N0_1 \ell^{\pm} \nu$) is almost the same as the
leptonic BR of W-boson for our considered parameter space. 

In this section, the sensitivity of the signal,Eq.~\ref{eq:sig} cross sections 
with the parameters are studied systematically by scanning those using 
{\tt NMSSMTools4.9.0}\cite{Ellwanger:2004xm} taking into account 
various constraints 
such as dark matter, flavor physics and direct searches at LEP and LHC experiments. 
In this numerical scan we use the following range of 
parameters:
\begin{gather}
\rm {0.1 < \lambda < 0.7;\ \ 0.1<\kappa < 0.7;\ \ 0< A_{\lambda}<2~TeV},
\ \ \rm {-9 < A_{\kappa} < -4~GeV}; \nonumber \\   
\rm {2 <\tan\beta<50; \ \ 140~GeV<\mu_{ eff}< 600~GeV} \nonumber \\
\rm {M_{Q_3}=M_{U_3}=1-3~TeV} ,\ \ {A_t = -3 ~-~ (+3) ~{ TeV}},
\label{eq:p1}
\end{gather}

The other soft masses are set as 
$$\rm {M_{Q_{1/2}} =M_{U_{1/2}} =M_{D_{1/2}}= M_{D_3}=M_{L_3}=M_{E_3} =A_{E_3}=1TeV}$$ 
$$\rm A_b=2{TeV}, M_{L_{1,2}}=M_{E_{1,2}}=200GeV, A_{E_{1,2}}=0.$$ 
%({\bf Scan is subject to Constraints})  

The important factors in this discussion are the mass and the 
composition of $A_1$ which is dominantly a singlet like.
In order to understand the variation of composition of $A_1$,
here we briefly revisit the Higgs mass matrix corresponding to CP-odd states. 
The initial 3$\times$3 CP odd Higgs mass matrix 
reduces to 2$\times$2 matrix after rotating away the 
Goldstone mode. 
Hence, the CP-odd mass matrix, $\rm M_P^2$,
in the basis of doublet(A) and singlet(S), 
is given by \cite{Ellwanger:2009dp,Miller:2003ay},

\bigskip
\begin{equation}
\rm  
M_P^2=\rm \left( \begin{array}{cc}
 M_A^2 & \lambda(A_{\lambda} -2 \kappa v_s )v   \\
        \lambda(A_{\lambda} -2 \kappa v_s )v    & M_S^2   \\
 \end{array} \right),
 \label{eq:a1mass}
 \end{equation}
 
where
\begin{equation}
\rm M_A^2 = \rm \frac{2 \mu_{eff}(A_{\lambda} +\kappa v_s)}{\sin 2 \beta}, ~~
\rm M_S^2 = \rm \lambda(A_{\lambda} +4 \kappa v_s)\frac{v_u v_d}{v_s} - 3 \kappa A_{k} v_s.
\label{eq:MA}
\end{equation}
This 2$\times$2 mass matrix can be 
diagonalized by an orthogonal rotation with an angle $\alpha$,  as 
given by,
\begin{equation} 
\rm \tan 2 \alpha = \rm \frac{2 M_{12}^2}{(M_A^2 -M_S^2)},
\label{eq:alpha}
\end{equation}
where $M_{12}^2 =\lambda (A_{\lambda} - 2 \kappa v_s)v$ and 
$v  = \sqrt{v_u^2 + v_d^2}$.
Obviously, two mass eigenstates $(A_1,A_2)$ are the mixtures of the doublet 
$(A)$ and the singlet $(S)$ weak eigen states.

$\bullet$ \underline{$\widetilde \chi_j^0 \rightarrow \widetilde \chi_1^0 A_1$, j=2,3}: 
The relevant part of the coupling (Higgsino-Higgsino-Singlet)
 for this decay channel is given by,
\br
g_{\tilde \chi_j^0 \tilde \chi_1^0 A_1} 	\approx \frac{i}{\sqrt 2} \lambda ~P_{13} \left(N_{j4} N_{13} + N_{j3} N_{14}  \right).
\label{eq:n2n1a1}
\er 
\begin{figure}[t]
\centering
\includegraphics[width=0.65\textwidth]{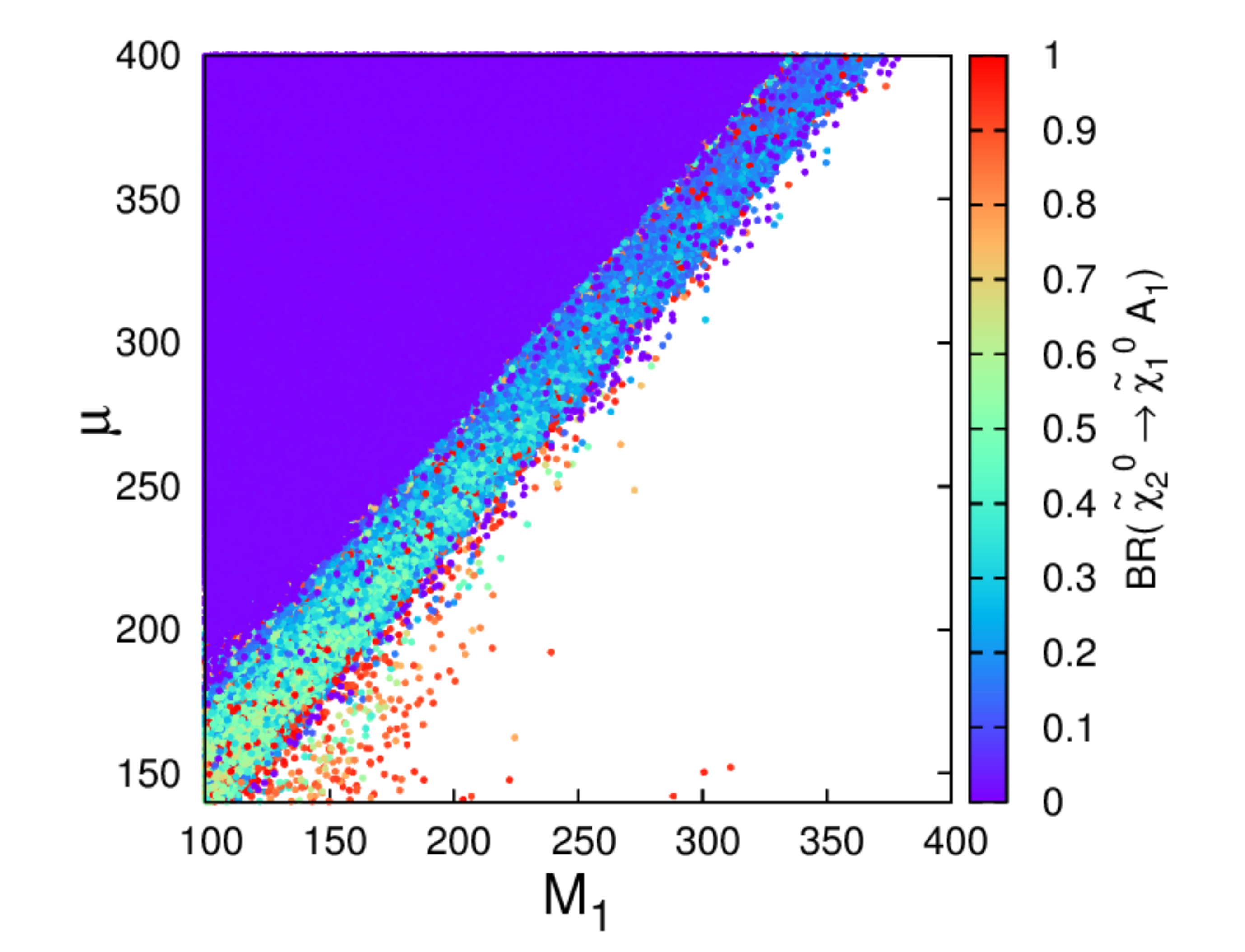}
\caption{\small BR($\N0_2 \to \N0_1 A_1)$ in the $M_1-\mu_{eff}$ plane. All energy units are in GeV.}
\label{fig:M1mu}
\end{figure}
Here $\rm P_{13}\sim \cos\alpha$ presents the singlino composition in $A_1$.
Hence, for very small values of $\sin\alpha$ this coupling favors only 
the Higgsino like $\N0_j$ and $\N0_1$ states.  
Note that, in the context of our signal, the gaugino like 
$\N0_j$ and $\N0_1$ states are not favoured in order to suppress 
the decay modes such as, 
$\N0_j \to \N0_1 Z, \ell \tilde \ell$.
This type of Higgsino like scenario can be achieved by 
setting $\mu_{eff} \sim M_1 < M_2$, which also makes 
$\tilde \chi_{2,3}^0$ and $\tilde \chi_{1}^0$ states almost degenerate, 
i.e $m_{\N0_2} \sim m_{\N0_1}$, a compressed like scenario. 
However, in order to have a reasonable sensitivity of this signal, the visible 
decay spectrum are expected to be little bit harder to pass 
kinematic thresholds, which 
can be ensured by setting the mass splitting, 
$\Delta m= m_{\tilde \chi_{2,3}^0}- 
m_{\tilde \chi_1^0}$ to a reasonable value. 
This requirement leads us to choose $M_1$ less than $\mu_{eff}$, 
but of course, not by a huge gap to retain sufficient Higgsino
component, making $\N0_1$ a bino-Higgsino mixed state.
In Fig.\ref{fig:M1mu}, we show the correlation of 
BR($\rm \N0_2 \to \N0_1 A_1$) in the $\rm M_1 - \mu_{eff}$ plane.
Notice that the 10\% or more BR($\rm \N0_2 \to \N0_1 A_1$) 
corresponds to the region 
$\rm M_1 \sim \rm \mu_{eff}$ and, we found that it remains to be valid 
for a wide range of $\lambda$ and $\kappa$.    
This figure clearly reflects the preferred choices of $M_1$ 
and $\rm \mu_{eff}$ for our considered signal channel.

{$\bullet$ $\underline{A_1 \to \gamma\gamma}$: The earlier studies 
\cite{Christensen:2013dra,Guchait:2015owa,Dermisek:2007yt,Arhrib:2006sx}
showed that the variation of BR of non SM-like NMSSM Higgs bosons in 
various decay channels 
is very dramatic depending on the region of parameters. 
For instance, the singlet like $A_1$ state decouples
from the fermions leading a suppression of the tree level decay modes 
$b\bar b$ and $\tau\tau$ and an enhancement of 
BR($A_1 \to \gamma\gamma$) channel
~\cite{Christensen:2013dra,Dermisek:2007yt,Arhrib:2006sx}.
The cause of having a finite partial $A_1 \to \gamma\gamma$ decay width
can be understood by examining the respective coupling  
structures of $A_1$ with two photons~\cite{Munir:2013wka}.
The $A_1$ state decays to two photons via loops  
comprising heavy fermions and 
charginos~\cite{Spira:1995rr,Spira:1997dg}, see Fig.\ref{fig:hgg}.
\begin{figure}[t]
\centering
  \includegraphics[height=4.0cm,width=17.0cm,trim={2cm 22cm 0 4.0cm},clip]{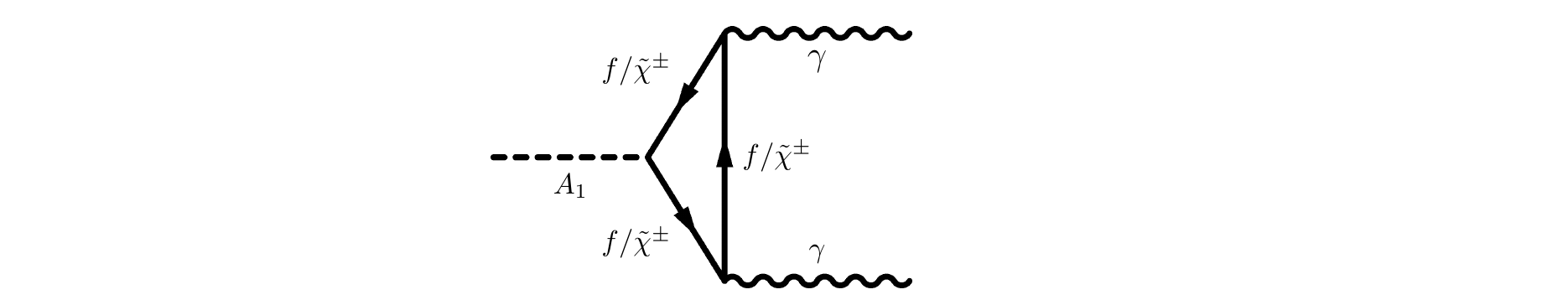}
\caption{\small Loop diagrams for the decay of $A_1$ to two photons,  
mediated by fermion $(f)$ and chargino($\tilde \chi^{\pm})$.}
\label{fig:hgg}
\end{figure}
The partial decay width of $A_1 \to \gamma\gamma$ can be obtained 
simply using the MSSM expression, but replacing the respective  
couplings to the NMSSM values. Thus, it is 
given as~\cite{Spira:1995rr,Spira:1997dg},
\begin{equation}
\Gamma(A_1 \rightarrow \gamma \gamma)= 
\frac{G_F \alpha_{em}^2 M_{A_1}^3}{32 \sqrt 2 \pi^3}
\left |\sum_f N_c ~e_f^2 ~g_f^{A_1}~ A_f(\tau_f) ~ + ~
\sum_{\tilde \chi^{\pm}_i} ~ g_{\tilde\chi^\pm_i}^{A_1} ~ A_{\tilde \chi^{\pm}_i}
(\tau_{\tilde \chi^{\pm}_i}) \right|^2.
\label{eq:a1gg}
\end{equation}
Here $\rm N_c$ is the QCD color factor, $e_f$ is the electric charge of the fermions $(f)$,
$A_x (\tau_x)$ are the loop functions given by,
\begin{equation}
A_{x} (\tau_x)=\tau_x \left(\sin^{-1} 
\frac{1}{\sqrt \tau_x} \right )^2,   \ \  \tau_x = \frac{4  M^2_{x}}{M_{A_1}^2};\ \  x = f,~\tilde\chi^\pm_i. 
\end{equation}
Here $g_f^{A_1}$ are the couplings of $A_1$ with the heavier 
fermions(f=top and bottom quarks), 
where as $g_{\tilde \chi^{\pm}}^{A_1}$ are the same with charginos, and all 
those are given by 
\cite{Ellwanger:2009dp},
\br
g_{u}^{A_1} = -i \frac{m_u}{\sqrt 2 v \sin \beta} P_{12}, \ \  
g_{d}^{A_1} =  i \frac{m_d}{\sqrt 2 v \cos \beta} P_{11}, 
\\
g_ {{\tilde \chi_i^{\pm}} 
{\tilde \chi_j^{\mp}} {A_1}} 	= \frac{i}{\sqrt 2} \left [ { \lambda} 
P_{13} U_{i2} V_{j2} - 
{g_2} (P_{12} U_{i1} V_{j2} +P_{11} U_{i2} V_{j1})  \right]
\label{eq:coup}
\er 
Here $P$ and (U,V) are the mixing matrices for pseudoscalar Higgs bosons
and chargino sector respectively and, in particular 
$P_{11}=\sin \alpha \sin \beta$ and $P_{12} =\sin \alpha \cos \beta$.
In the pure singlet limit of $A_1 ( P_{11},P_{12} \sim 0$),
see Eq.~\ref{eq:ddx} and, hence  
the fermion couplings($g_u^{A_1},g_d^{A_1}$) approach to almost 
negligible value($\sim 10^{-5}$),
%\br
%g_{u}^{A_1}, g_{d}^{A_1}  \xrightarrow{singlet~ A_1} ~ \sim 10^{-5},
%\label{eq:singlet}
%\er 
and, hence the corresponding fermionic loop contribution 
in Eq.~\ref{eq:a1gg} are extremely suppressed. On the other hand,  
the presence of Higgsino composition in the chargino state yields a 
favorable coupling with $A_1$ via the singlet-Higgsino-Higgsino 
interaction(see the term proportional to $\lambda$ in Eq.\ref{eq:coup}).
Needless to say, that it is purely a typical NMSSM effect.
Naturally, it is interesting to identify the region of the parameter space 
which offers a finite partial width of $A_1 \to \gamma\gamma$ mode.
We try to study it by examining the mixing of 
CP odd Higgs bosons states via the mass matrix, 
Eq.~\ref{eq:a1mass} ~\ref{eq:MA}.
Recall, that a very small value of $\sin\alpha$ leads a singlet 
dominated $A_1$ state resulting a 
suppression of its couplings with the fermions. 
Following the mass matrix, 
it can be realized 
very easily that the lighter CP odd state $A_1$, 
can be a very much singlet like in the presence of negligible mixing 
between A and S states and, 
essentially it can happen due to either of the following two conditions: 
\begin{enumerate}
\item
$M_A^2 >> M_S^2, M_{12}^2$ \emph{i.e} the heavier state is too heavy and
purely doublet like where as the lighter state is singlet, 
a decoupled type of scenario.  
\item
$M_{12}^2=(A_{\lambda}- 2 \kappa v_s) \sim 0$, \emph{i.e}, 
a cancellation between two the terms in the off-diagonal element.
\end{enumerate}
These two scenarios are illustrated in Fig.\ref{fig:mam12}, 
presenting the range of $\rm M_A^2$ and $\rm M_{12}^2$ 
(Eq.\ref{eq:a1mass},~\ref{eq:MA}), 
corresponding to BR($A_1 \to \gamma\gamma) \gtrsim 10 \%$.  
\begin{figure}[t]
\hspace{-1.5cm}
\includegraphics[height=7.0cm,width=10.0cm]{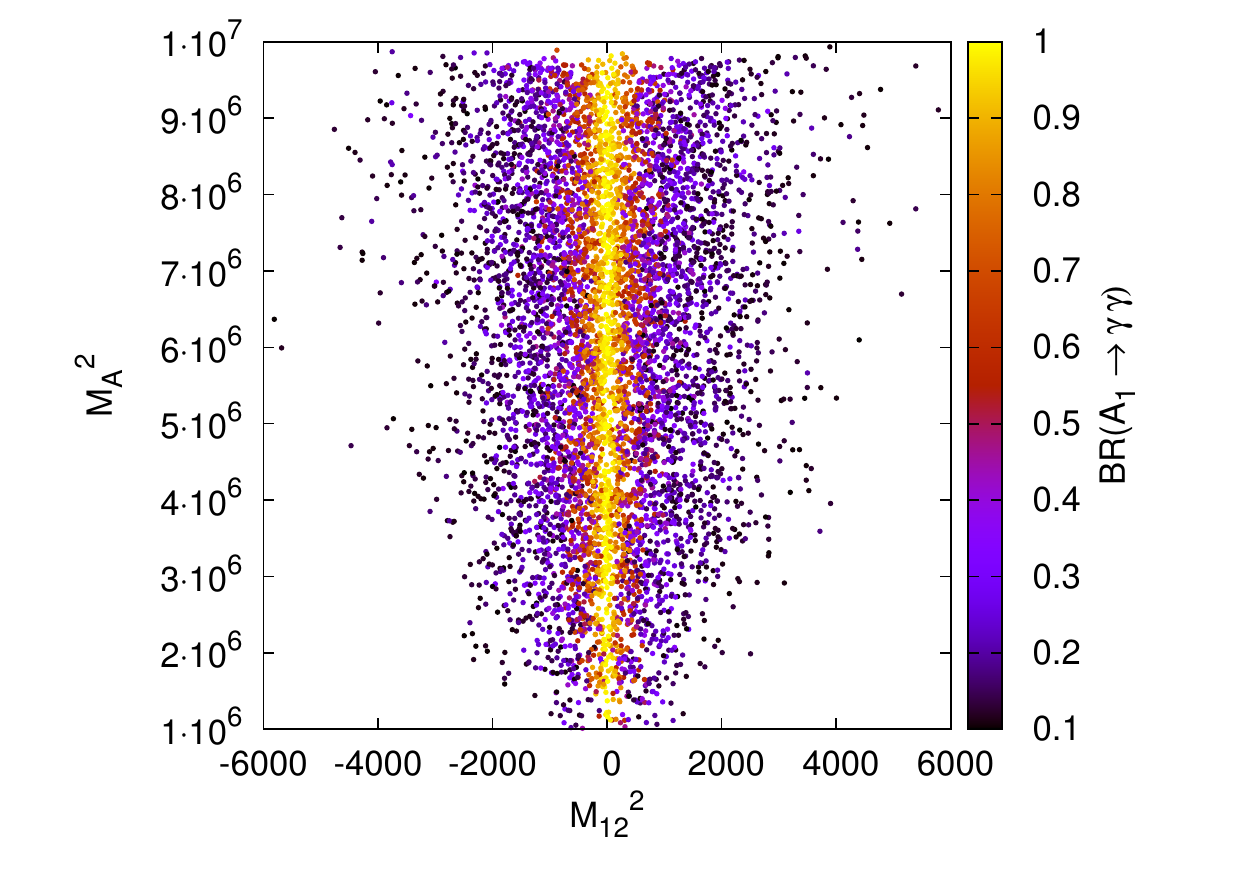}
\hspace{-1.4cm}
\vspace{-0.2cm}
\includegraphics[height=6.8cm,width=10.0cm]{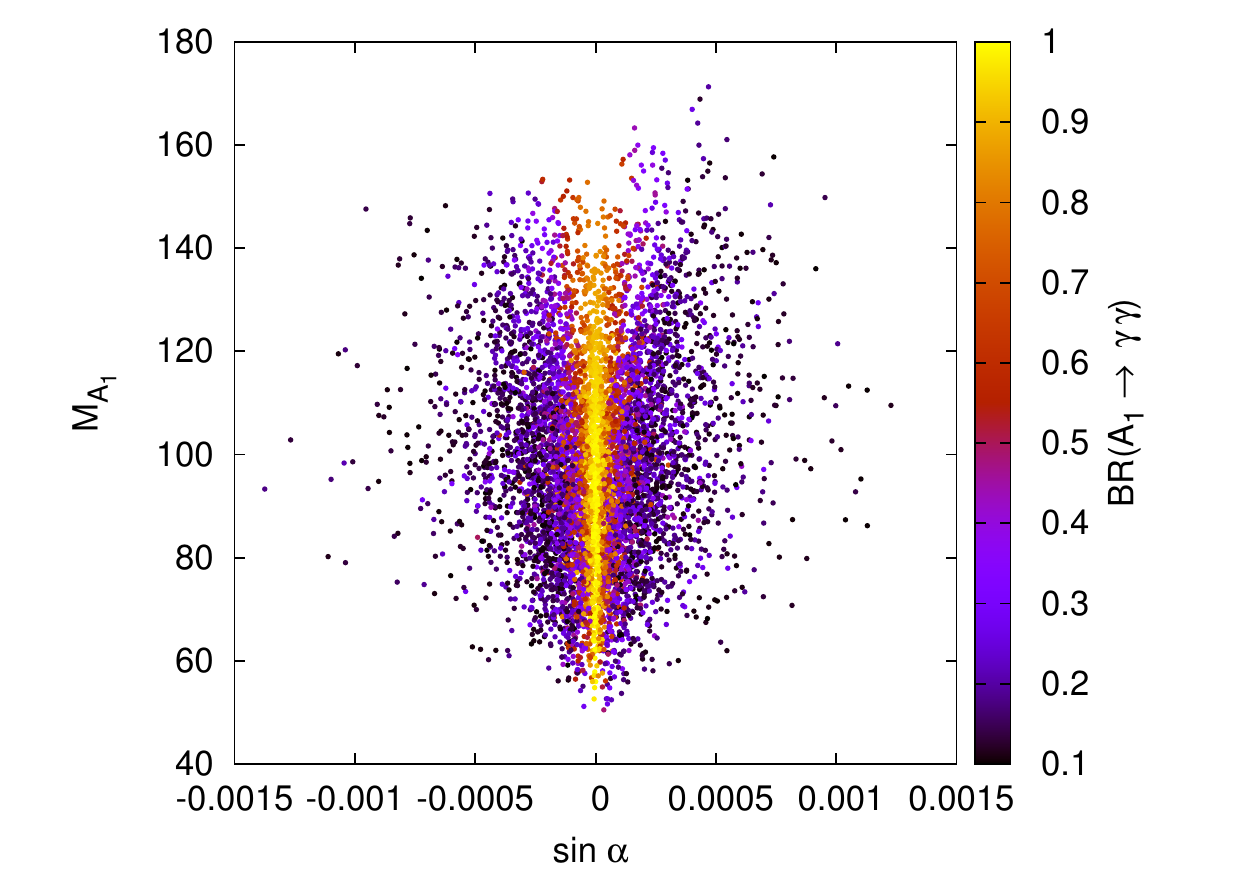}
\caption{\small BR($A_1 \rightarrow \gamma \gamma$) in the $M_A^2-M_{12}^2$ (left) and $\sin\alpha - M_{A_1}$ plane (right). The other parameters are varied for the range, as given in Eq.\ref{eq:p1}. All energy unit are in GeV.}
\label{fig:mam12}
\end{figure}
In the left panel, we present the range of diagonal term $\rm M_A^2$ and 
the off-diagonal element $M_{12}^2$ of the mass matrix $\rm M_P^2$, 
Eq.~\ref{eq:a1mass}.
As expected, for very low values of 
$\rm M_{12}^2 (\sim 0)$ and corresponding to larger 
values of $\rm M_A^2 \sim 10^6$,
BR$(A_1 \to \gamma \gamma)$ appears to be ($\gtrsim$80\%), and even for 
the case $0 < |M_{12}^2| << M_A^2 $, it can be about 10-20\%.  
It also indicates that the 
BR($A_1 \to \gamma\gamma)$ becomes almost $100 \%$ for the 
scenario $M_{12}^2 \sim 0$,
\emph{i.e}  $A_{\lambda} \sim 2\kappa v_s$. 
Moreover, we show the range of mixing angle in terms of 
$\sin \alpha$ and the mass of $A_1$ 
in Fig.\ref{fig:mam12}(right), corresponding to the range 
of $M_{12}^2$ and $M_A^2$, as shown in the left panel of the same figure. 
It clearly confirms the smallness of the mixing angle 
responsible to yield a large BR($A_1 \to \gamma\gamma)$ and 
it occurs for a wide range of $\rm M_{A_{1}}$.
Similarly,
corresponding to the range of parameters as shown in Fig.\ref{fig:mam12}, for
which BR($A_1 \to \gamma\gamma \gtrsim$10\%), 
the relevant range of $A_{\lambda}$ and $\mu_{eff}$ are shown 
in the $\lambda-\kappa$ plane in the left and right panel
of Fig.~\ref{fig:alka} respectively.
\begin{figure}[t]
\hspace{-1.5cm}
\includegraphics[height=7.0cm,width=10.0cm]{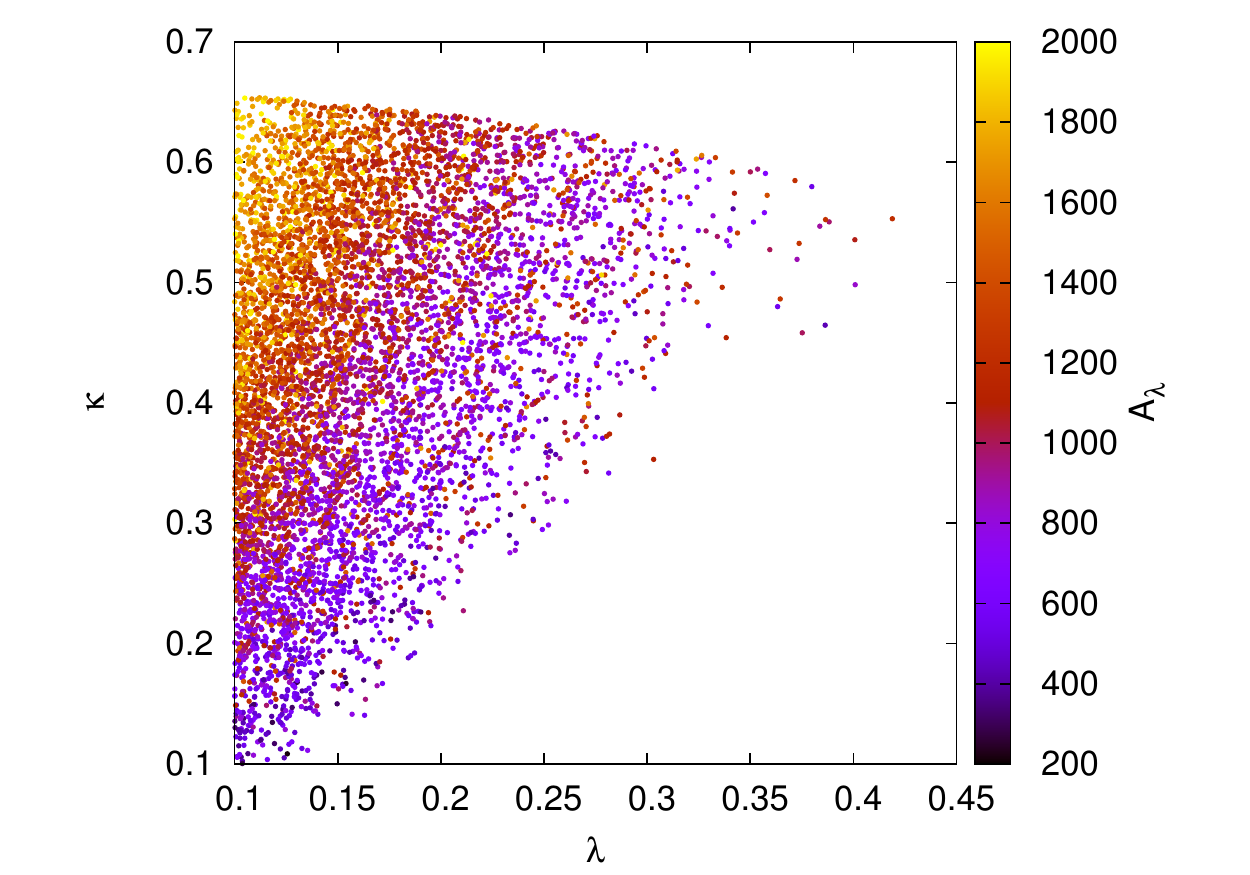}
\hspace{-1.3cm}
\vspace{-0.1cm}
\includegraphics[height=7.0cm,width=10.0cm]{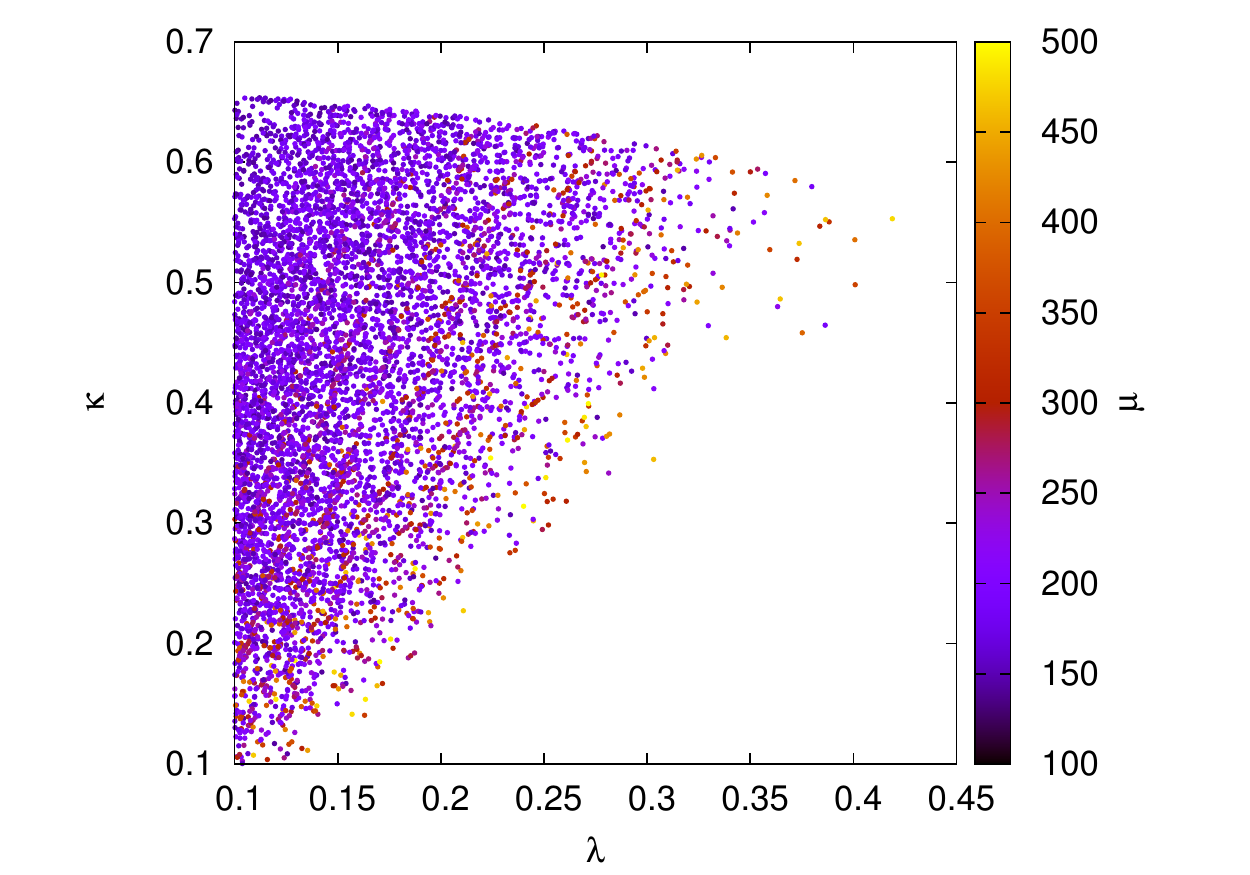}
\caption{\small $BR(A_1 \rightarrow \gamma \gamma$)($\ge 10 \%$) in 
the $\lambda-\kappa$ plane for the range of $A_\lambda$(left) 
and $\mu_{eff}$(right). The other parameters are varied for the range, 
as given in Eq.\ref{eq:p1}.}
\label{fig:alka}
\end{figure}
It is observed that a reasonable wide ranges of $\lambda$(0.1 -- 0.4) 
and $\kappa$(0.1 -- 0.65)
can provide a large 
BR($A_1 \to \gamma\gamma$) for a larger range 
of $A_{\lambda}$ and for a moderately large values of $\mu_{eff}$.
It is to be noted also that preferably Higgsino like lighter chargino  
i.e a smaller $\mu_{eff}$ as compared to $M_2$, required in order to 
enhance the partial width of this channel.
    
Finally, based on the above observations about the parameter dependence 
of the production 
cross sections, BR($\N0_2 \to \N0_1 A_1$) and BR($A_1 \to \gamma\gamma$),
we set up few benchmark points (BP) in order to present results.
In summary, the preferred choices are, $\N0_1$ as a bino-Higgsino mixed state,
$\N0_{2,3}$ and $\C1$ primarily Higgsino like, i.e $\rm M_1 < \mu_{eff}$, 
but not 
with large gap between $M_1$ and  $\mu_{eff}$, and $\rm M_2$ set 
to a larger value satisfying  
$M_2 > \mu_{eff}$. In Table \ref{tab:spectra}, we show six BPs and presenting
the corresponding parameters, masses of relevant particles and BRs. 
Notice that BP1-BP4 present comparatively lighter masses of 
chargino and neutralino 
states, whereas these are massive for BP5 and BP6. 
The values of $M_{A_1}$ are chosen
in such a way that the decay of the SM Higgs to a pair of $A_1$ is 
forbidden in order to make it compatible with recent  
SM Higgs boson results~\cite{Khachatryan:2016vau}.
For all BPs, the lightest CP even Higgs boson, $H_1$ is SM-like.
Although, both the $\N0_2$ and $\N0_3$ neutralino states are 
Higgsino like, but, more precisely, the coupling strength depends on the 
kind of Higgsino composition, either it is 
$\rm \tilde H_u$ or $\rm \tilde H_d$ 
(see Eq.~\ref{eq:n2n1a1}) like. Notice that, for BP4, because of the higher
mass of $A_1$, the $A_1 \to Z\gamma$ also opens up and found to be its BR 
around $\sim$ 2\%. This decay channel of $A_1$ can give rise to a 
spectacular signal with the final state $\rm Z\gamma$ along with a lepton and
$\PMET$, when it is produced through the production mechanism, as shown in 
Eq.\ref{eq:1}.
\begin{table}[t]
\begin{center}
{\renewcommand{\arraystretch}{1.0}
\begin{tabular}{|c|c|c|c|c|c|c|c|  }
\hline
 & BP1&BP2 &BP3& BP4 &BP5& BP6  \\ \hline
\hline
$\lambda$ &0.29&0.40&0.10&0.53&0.64&0.50     \\
$\kappa$ &~0.37 &0.45  &0.20 &0.39 &0.36&0.48     \\
$ \tan \beta$ &6.46 &6.46&11.0&4.0&2.5&2.84\\
$M_{A}$ &1722&340.7&1311.5&1262.4&1436.9&1655.8 \\
$A_{\kappa}$ &-4.97 & -4.97 &-3.9 &-5.8 & -6.5&-9.37 \\
$\mu_{eff}$ &342.4 &200.0 &158.5  &365.4 &636.8 &540.7\\
$M_1$ &300&150.0&135.4 &275.9& 605.8 &514.0\\
$M_2$ &606.6&606.6&1000.0 &9000&1857.4 &1597.1\\
%$M_{Q_3},~M_{U_{3}},~M_{D_3}$&2019,~ 2802,~1000&2019,~ 2802,~1000&2019,~ 2802,~2000   & 2019,~ 2802,~2000 &   \\
%$A_t,~A_b$   &3765,~ 2000 &3765,~ 2000& 3765,~ 2000 &3765,~ 2000   &   \\ 
\hline \hline
$M_{\tilde \chi_1^0}$&280.6 &131.4 &113.4 &261.8 &578.3&488.5  \\
$M_{\tilde \chi_2^0}$&356.4 &210.0 &169.0 &379.1 &657.5&559.8  \\
$M_{\tilde \chi_3^0}$&356.7 &215.6 &182.3 &385.5 &661.0&572.7  \\
$M_{\tilde \chi_1^+}$&340.0 &199.3 &161.7 &377.5 &648.6&550.6 \\
$M_{A_1}$           &62     &76    &63.1  &105.2 &62.8&66.8 \\
$M_{H_1}$           &124& 124 & 124&124 &125 &123 \\ \hline \hline
$BR( \chi_2^0 \rightarrow \tilde \chi_1^0 A_1)$ &0.92 &0.83  &0.0  &0.44
&0.98&0.05  \\
$BR( \chi_3^0 \rightarrow \tilde \chi_1^0 A_1)$ &0.27  &0.31  &0.52 &0.002
&0.11&0.97  \\
$BR(A_1\rightarrow \gamma\gamma)$  &0.79  &0.91  &0.98 &0.87 &0.97& 0.97  \\
%$\Gamma(A_1\rightarrow \gamma\gamma)$ &2.8E-9&2.5E-8&1.3E-9&3.6E-8
%&3.4E-9 &3.5E-9  \\
\hline
\end{tabular}
}
\caption{\small Parameters, masses, and BRs for six benchmark points.}
\label{tab:spectra}
\end{center}
\end{table}

%%%%%%%%%%%%%%%%%%%%%%%%%%%%%%%%%%%%%%%%%%%%%%%%%%%%%%%%%%%%%%%%%%%%%%%%%%%
\normalsize
%%%%%%%%%%%%%%%%%%%%%%%%%%%%%%%%%%%%%%%%%%%%%%%%%%%%%%%%%%%%%%
\section {Signal and Background}
In this section we present the detection prospect of finding the signal
$\gamma\gamma +\ell^{\pm} + \slashed E_T$ 
at the LHC with the center of mass energy, $\sqrt S$= 13 TeV, corresponding 
to a few integrated luminosity options. 
As mentioned in the previous section, the signal events
appear from both the $\C1\N0_{2}$ and $\C1\N0_{3}$ production 
following the cascade decays, $\N0_{2,3}\to \N0_1 A_1$, 
and $A_1 \to \gamma\gamma$(Eq.~\ref{eq:1}). The lepton originates mainly from 
$\tilde\chi_1^\pm \to \ell^{\pm} \nu\N0_1$ decay and the missing transverse 
energy ($\slashed E_T$) arises due to the presence of massive LSPs,
in addition to almost massless neutrinos. 
The dominant SM background contributions come from the following 
processes,
\br
pp  \to { W\gamma, \ \ Z\gamma, \ \ W\gamma\gamma, \ \  Z\gamma\gamma},
\er
with the leptonic decays of $W/Z$. 
Note that in the first two cases, the second photon originates primarily  
from the initial state, radiated by incoming quarks. In addition, the another
potential source of backgrounds are due to the faking of jets as photon 
in the process, 
\br
pp \to W \gamma j, \ \ Z \gamma j,
\er   
and interestingly, it is  found to be the dominant ones.  
 
In our simulation we generate signal events using 
{\tt PYTHIA6}~\cite{Sjostrand:2006za}
providing spectrum of SUSY particles and BR of various decay channels
through SLHA file\cite{Skands:2003cj}, obtained from
{\tt NMSSMTools}~\cite{Ellwanger:2004xm}, 
corresponding to our chosen parameter space, 
as shown in Table~\ref{tab:spectra}.
The background events with 2-body at the final state($W\gamma,, Z\gamma$)
are generated directly using  
{\tt PYTHIA6}, while processes consisting 3-body are simulated using 
the MadGraph~\cite{Aad:2015jqa} and then {\tt PYTHIA6 } is used  
for showering.
The generated events are
stored in the standard HEP format (STDHEP) \cite{stdhep} 
to pass them through 
{\tt Delphes3.2.0}~\cite{deFavereau:2013fsa}
to take into account the detector effects.
In our analysis we have used the default CMS card in Delphes,
but results are also checked with ATLAS default card  
and not much differences are observed.

The objects in the final state such as, electron, photon and 
missing transverse energy are identified and reconstructed using Delphes 
based algorithms \cite{deFavereau:2013fsa}. However, 
for the sake of completeness,
we describe very briefly the object reconstruction 
techniques followed in the Delphes. 
\begin{itemize}
\item
Lepton Selection: 
The electrons are reconstructed using the information from the tracker and
ECAL parameterizing the combined reconstruction efficiency as a function
of the energy and pseudorapity. The muons are reconstructed using the 
predefined reconstruction efficiency and the final momentum is obtained
by a Gaussian smearing of the initial 4-momentum vector. In our simulation, 
both the electrons and the muons are selected, imposing cuts on the 
transverse momenta
($p_T^\ell$) and pseudo rapidity ($\eta^l$) of lepton as,
\br
p_{T}^{\ell} \ge 20 { ~GeV}; ~|\eta^l| \le 2.5; \ \  (l = e, \mu),
\label{eq:lcut}
\er
where $\eta^\ell$ restriction is due to the limited tracker coverage. 
The leptons are required to be isolated by demanding the total transverse
energy $E_{T}^{ac}(\ell) \le 20 \%$ of the $p_T^\ell$,
where $E_{T}^{ac}(\ell)$ is the scalar sum of transverse energies of particles
with minimum transverse momentum 0.5 GeV
around the lepton direction within a cone size of $\Delta R=0.5$.
\item
Photon Selection: The genuine photons and electrons that
reach to the ECAL having no reconstructed tracks are considered as 
photons in the Delphes 
neglecting the conversions of photons into electron-positron pairs.
In the present version of Delphes~3.2.0, the fake rate of photons are not
simulated. In our simulation, we select photons subject to cuts,
\br
p_T^\gamma > 20 ~GeV; \ \ |\eta^{\gamma}|<2.4,
\er 
but excluding the $\eta$ region, $1.44<|\eta|^{\gamma} < 1.57$.
The isolation of photon is ensured by measuring the sum of transverse momenta
$E_T^{ac}(\gamma)$ of all particles around $\Delta R$=0.5 along the  
of the axis of the photon and transverse momentum more than 0.5 GeV. 
We consider photon is isolated if,
\br
E^T_{AC}(\gamma)<0.2 ~p_T^{\gamma}.
\er
\item
Missing transverse energy: 
In the Delphes, the missing transverse energy is estimated from
the transverse component of the total energy deposited in the detector, 
as defined,
\br
\vec E{\!\!\!/}_T = - \sum {\vec p_T(i)}
\er
where $i$ runs over all measured collection from the Detector. In the signal 
event $\slashed E_T$ is expected to be harder as it appears  
due to the comparatively heavier object $\N0_1$, where as in the SM 
it is mainly due to the neutrinos. Hence, $\slashed E_T$ may be a
useful variable to isolate background events by a good fraction without
affecting signal events too much.
A cut,
\br
\slashed E_T > 50 ~GeV,
\label{eq:metcut}
\er 
is applied in our simulation and observed that 
a substantial fraction ($\gtrsim 50$\%) of background events are 
rejected with a mild loss of signal events.
\end{itemize}
With a goal to separate out the signal from the background events,
we investigate several kinematic variables.
We notice that the $p_T^\gamma$ are comparatively 
harder in the signal than the background events. This can be 
attributed to the fact
that the photons in the signal events originate from  $A_1$ decay,
which is to some extent expected to be boosted as it is produced from
heavier neutralino states. On the other hand, in the background process 
photons arise due to soft or hard emission accompanied 
with a $W/Z$ boson and are not as boosted as in the
signal events. Hence, we impose following hard cut on the leading($\gamma_1$) 
photon and little mild on the sub-leading ($\gamma_2$) photon to eliminate  
background events,
\br
p_T^{\gamma_1}>40~ GeV; \ \ p_T^{\gamma_2}> 20~ GeV.
\label{eq:gammaptcut}
\er 
Moreover, interestingly, we observed that the distribution of $\Delta R_{\gamma_1\gamma_2}$,
defined as, 
\br 
\Delta R _{\gamma_1 \gamma_2} = \sqrt{(\eta_{\gamma_1} - \eta_{\gamma_2})^2
+ (\phi_{\gamma_1} - \phi_{\gamma_2})^2}  ~,
\er 
presents a characteristic feature for the signal events.
Two photons in signal events originating from a
comparatively massive $A_1$ are expected to be correlated and appear 
without much angular separation between them, unlike the background events,
where these are not directly correlated and come out with a comparatively 
wider angular separation. This interesting feature is clearly demonstrated 
in the distribution of $\Delta R_{\gamma_1\gamma_2}$, as shown in 
Figure~\ref{fig:drg1g2-philg2} (left), for both the signal and 
dominant backgrounds, such as $W \gamma,~ W \gamma \gamma,~ W \gamma j$. Note that, $\Delta R_{\gamma_1\gamma_2}$
distributions are subject to cuts given 
by Eqs.~\ref{eq:lcut}, \ref{eq:metcut}, \ref{eq:gammaptcut}.
%%%%%%%%%%%%%%%%%%%%%%%%%%%%
\begin{figure}[t]
\hspace{-0.3cm}
\includegraphics[height=8.0cm,width=8.0cm]{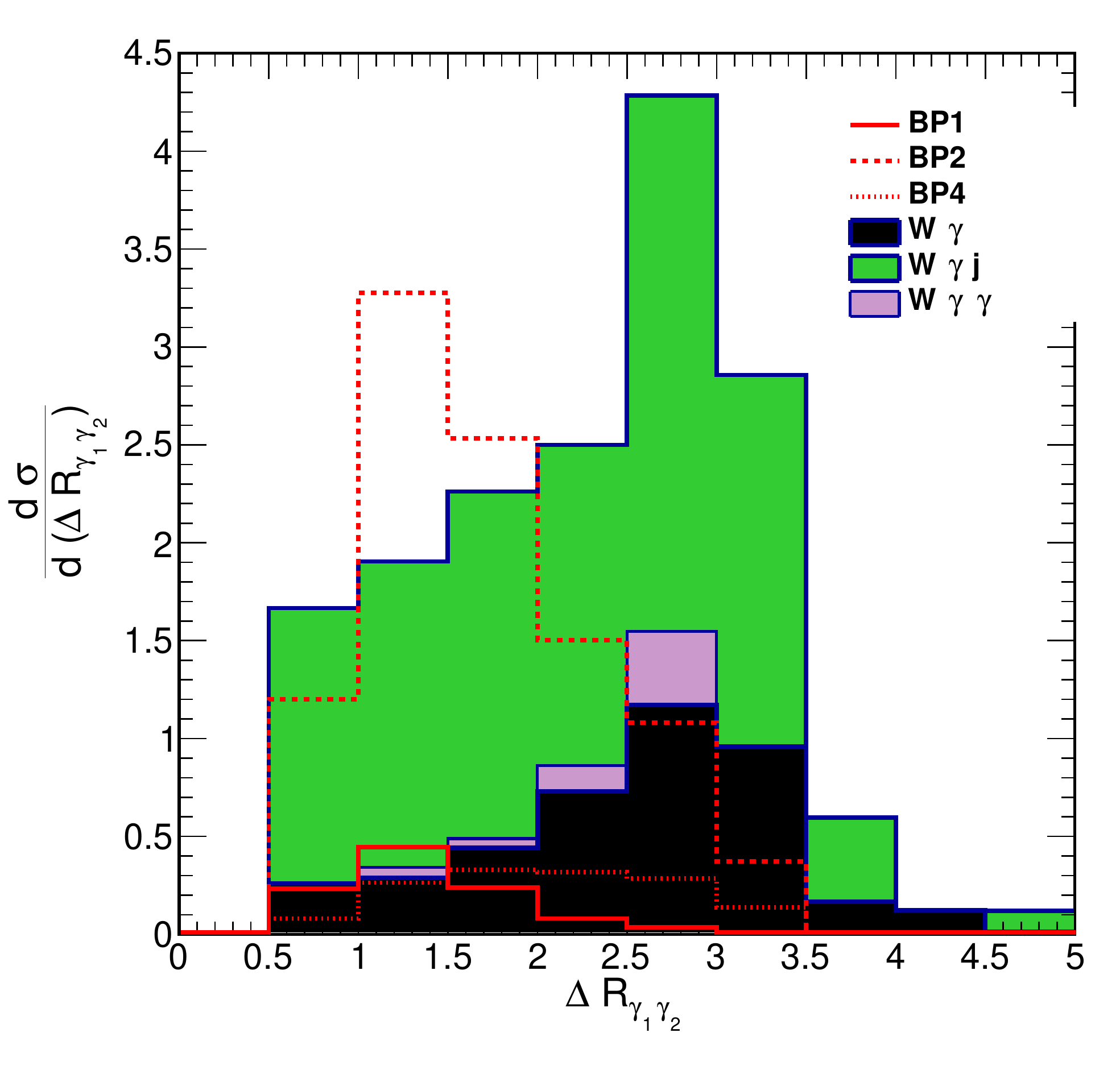}
\hspace{-0.1cm}
\vspace{-0.4cm}
\includegraphics[height=8.0cm,width=8.0cm]{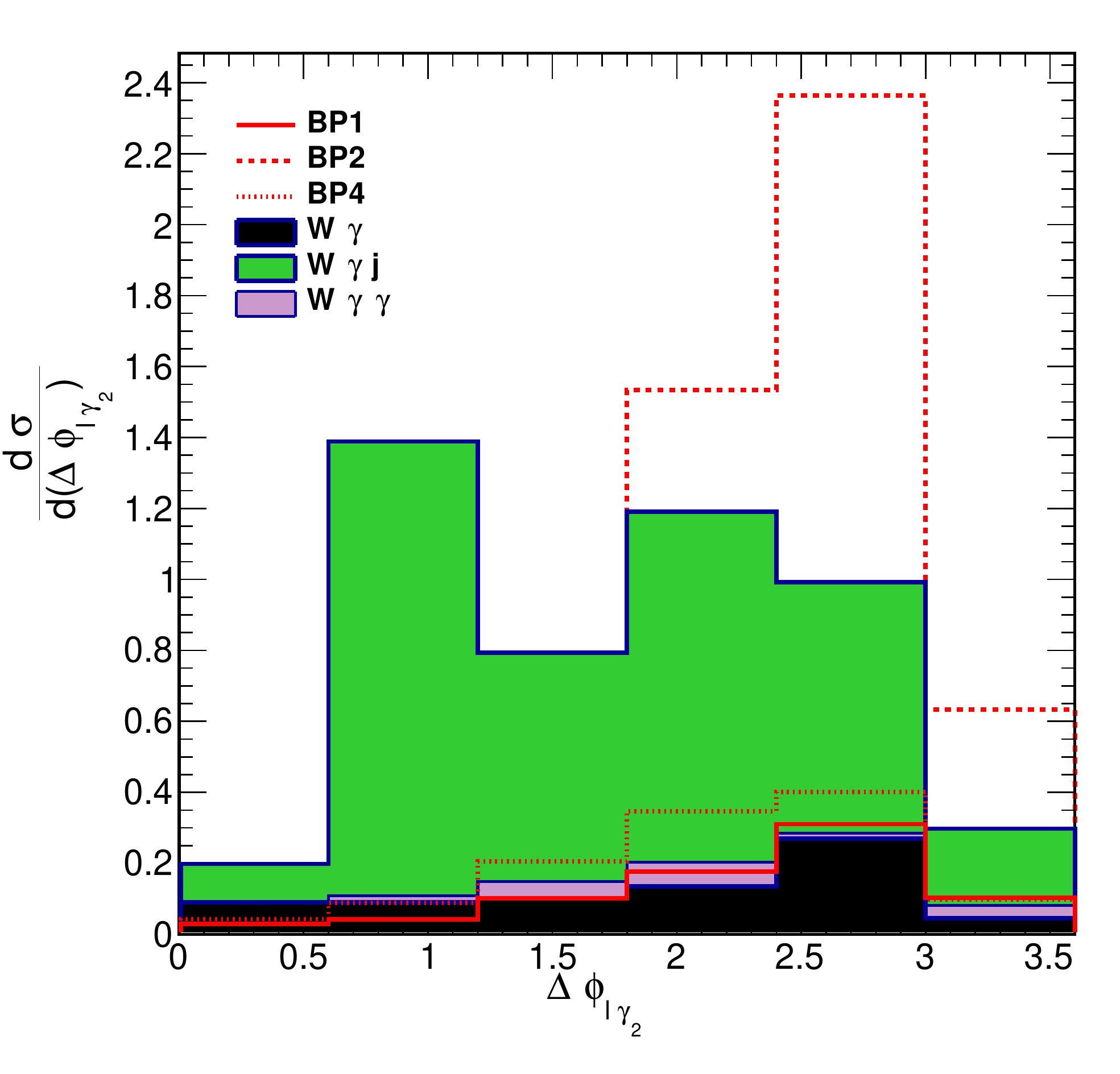}
\caption {\small $\Delta R_{\gamma_1\gamma_2}$(left) 
and $\Delta \phi_{l \gamma_2}$(right) distribution for both the signal 
and dominant backgrounds. These are subject to selection cuts,
Eqs.~\ref{eq:lcut}, \ref{eq:metcut}, \ref{eq:gammaptcut}.}
\label{fig:drg1g2-philg2}
\end{figure}
It displays a clear difference, where the signal events are distributed in 
the lower region of $\Delta R_{\gamma_1\gamma_2} $},
where as the background events mostly appear towards the higher side. 
Evidently, this characteristic feature  
can be exploited to improve the purity of the signal events. 
Optimizing the selection of $\Delta R_{\gamma_1 \gamma_2}$, we require,
\br
\Delta R _{\gamma_1 \gamma_2}  \le 2.0
\er 
in our simulation and eliminate a good fraction of 
background events. Finally, to minimize the background contamination further, 
in particular due to the most dominant $W\gamma j$ process, we construct 
another observable, the difference in the azimuthal angle between 
the lepton and the sub-leading
photon \emph{i.e} $\Delta\phi_{\ell\gamma_2}$. 
In Fig.~\ref{fig:drg1g2-philg2}(right), we present the distribution
of $\Delta\phi_{\ell\gamma_2}$ for both the signal and the dominant 
backgrounds($\rm W\gamma, W\gamma j, W\gamma\gamma$).        
This distribution clearly shows a difference in behavior of the signal events
which are distributed towards the higher values of 
$\Delta \phi_{\ell\gamma_2}$, while the dominant $W\gamma j$ background 
does not show any such pattern. 
Hence, a selection of $\Delta\phi_{\ell\gamma_2}$ as,
\br
\Delta\phi_{\ell\gamma_2}> 1.5.
\er   
further suppresses the $W\gamma j$ background without much reduction of 
the signal size. Also note that in this selected region
of $\Delta \phi_{l\gamma_2}$, 
only the signal contribution corresponding to the BP1 point is large, 
while for the other BPs, it is more or less at the same level 
as backgrounds.  
Implementing all selection cuts together in the simulation, we 
achieve a reasonable signal
sensitivity as discussed in the next section.

\section{Results}
%%%%%%%%%%% Put table %%%%%%%%%%%%%%%%%%%%%%%%%%%%%%%%%%%%
\begin{table}[tb]
\centering
{\renewcommand{\arraystretch}{0.9}
\begin{tabular}{|l|l|l|l|l|l|l|l|l|l|}
\hline
&Process& $\sigma$(NLO) &$N_{ev}$ &$N_{\gamma} \ge 2$ &$ N_l=1$
&$\slashed E_T \ge 50$ & $\Delta R_{\gamma_1 \gamma_2} $ \hspace{-0.3cm} 
&$\Delta \phi_{l \gamma_2}$&  $\sigma \times \epsilon$ \\ 
&&&&&&& $\le 2$&$\ge 1.5$ & (fb)\\ \hline \hline
BP1& $\N0_2 \C1$       &36.4 fb &0.3L &7124 &886 &569 &502 &426 &0.38    \\
& $\N0_3 \C1$          &44.8 fb &0.3L &7006 &879 &587 &519 &431 & 0.14   \\
BP2&$\N0_2 \C1$        &335 fb &0.3L&9303 &1140 &590 &415 &346 & 2.9   \\
& $\N0_3 \C1$          &442 fb&0.3L &9593 &1213 &682 &499 &418 & 1.7   \\
BP3&$\N0_3 \C1$        &539 fb &0.3L&5755 &589 &312 &270 &240 & 2.2  \\
BP4&$\N0_2 \C1$        &61.1 fb &0.3L&14750 &2555 &1916 &910 &738 & 0.6     \\
& $\N0_3 \C1$          &43.9 fb &0.3L&14827 &2447 &1873 &935 &730 & 0.002  \\
BP5&$\N0_2 \C1$        &4.00 fb &0.3L&7798 &1023 &715 &598 &475 &0.060  \\
& $\N0_3 \C1$          &1.80 fb &0.3L&8292 &1111 &809 &694 &540 & 0.003  \\
BP6& $\N0_2 \C1$       &8.80 fb   &0.3L&7549  &893  &497 &353 &288 & 0.004 \\
& $\N0_3 \C1$          &4.90 fb &0.3L&9135  &1132 &813 &634 &517 & 0.080\\ \hline \hline
&$W \gamma$        &215 pb  &30M &15002 &1117 &272 &65 &47 & 0.33     \\
&Z $\gamma $       &103 pb &30M &14792 &1506 &52 &12 &10 & 0.03 \\

Bkg.&$W \gamma$ j      & 125 pb  &2.1M  &2987 &282 &137 &49 &30 & 1.80   \\
&Z$\gamma$ j       & 45   pb     &2.1M  &2531 &1203&27  &10 &6  & 0.13   \\
&W $\gamma \gamma$ & 407 fb  &0.5L &6011 &760 &260 &66 &47 & 0.40   \\
&Z $\gamma\gamma$  & 257 fb  &0.5L &5312 &233 &12 &7 &4 & 0.02  \\
\hline
\end{tabular}
\caption{\small Event summary for the signal and 
backgrounds(Bkg) subject to a set of cuts. The last column presents the cross section after multiplying the acceptance efficiency including BRs.}
\label{tab:event}
}
\end{table}
In Table~\ref{tab:event}, we present the summary of our simulation 
for both the signal and the SM backgrounds showing the number of events 
remaining after applying a given set of cuts. 
The results are shown for the signal corresponding to six BPs
as shown in the Table~\ref{tab:spectra}. 
The third column presents the production cross sections
and $N_{ev}$ in the 4th column indicates the number of events simulated 
for each processes. 
A k-factor 1.3 is used for the signal cross section in order to 
take into account NLO effects \cite{Beenakker:1999xh}.
The NLO cross sections for background processes are evaluated using 
${\tt MadGraph~{aMC}@NLO}$ \cite{Alwall:2014hca} subject to 
$p_T^\gamma > 10~GeV$ and $|\eta^{\gamma}|<2.5$ for photons,
where as $p_T^j> 20~GeV$ and $|\eta^j|< 5 $ are also used for 
accompanied jets at the generating level. 
Requirement of two hard photons and single lepton 
reduce the background contributions substantially by 3-5 orders of 
magnitude, where as the signal events decrease by about an order. 
The $\slashed E_T >50~GeV$ selection
is very effective in 
suppressing backgrounds, in particular process accompanying 
with a $Z$-boson in 
which case there is no genuine source of $\slashed E_T$. The 
selection of $\Delta R_{\gamma_1\gamma_2}$ appears to be very useful, 
as discussed 
above, in eliminating backgrounds by 60-80\% with a marginal reduction 
in signal events. 
Evidently, the dominant background contamination turn out to be due to 
the $W\gamma j$, which is about 65\% of the total background contribution. 
Notably, the background processes associated with a $Z$ boson 
are not contributing significantly, because of the
requirement of single lepton and a strong $\slashed E_T$. The signal 
benchmark points BP2 and BP3, comparatively with lower masses of 
$\C1$ and $\N0_2$ yield larger event rates, primarily due to the large 
production cross sections. The last columns shows the cross sections 
normalized by the selection efficiency due to set selections 
for each processes and parameters.   
space.  
\begin{table}[t]
\centering
{\renewcommand{\arraystretch}{0.8}
\begin{tabular}{ | l | l | l | l |l| l| l| l|l|l|l|l|l|l|l|l|}
\hline
Process  &BP1 &BP2 &BP3 & BP4 & BP5 &BP6 \\ \hline
$\sigma\times\epsilon$~(fb) & 0.52
& 4.6 & 2.2  & 0.6 & 0.063 & 0.084\\
\hline
${\cal L} ~{\rm (fb^{-1})}$ & \multicolumn{6}{|c|}{$S/\sqrt{B}$} \\ \hline
100
&3.1 & 28.1 & 13.3
&3.5 & 0.40  & 0.50 \\ 
300 
&5.4 & 48.7  & 23.9
&6.0 & 0.67 & 0.88 \\ 
1000
&9.8 & 89.0 & 42.0 & 11.0 & 1.22 & 1.60    \\ \hline
\end{tabular}
\caption{\small The signal cross sections after multiplying the 
acceptance efficiency
including BRs(2nd row) and significance $(S/\sqrt{B})$ 
for three integrated luminosity options $100, 300$ and $1000$ $\rm fb^{-1}$. 
The total background cross-section is 2.74~fb. }
\label{tab:sb}
}
\end{table}

In Table~\ref{tab:sb}, we show the sensitivity of the signal presenting
the significances ($S/\sqrt{B}$) for three integrated
luminosity options $100$, $300$ and $1000 \rm~\invfb$. The total 
background cross section is estimated to be about 2.74~fb. 
In this table the second row presents the signal cross section 
corresponding to each BPs.
The significances are quite encouraging for the lower 
masses ($\le 400~GeV$) of $\C1,~\N0_{2,3}$ and
for $A_1$ $\sim$ 60 -100 GeV, even for low integrated luminosity ${\cal L}=
$100$\invfb$. However, for the higher range of masses 
(BP5 and BP6), the sensitivity is very 
poor due to tiny production cross sections. 
We emphasize again that in order to obtain a sizeable signal rate, 
the chosen parameter space are happen to be a compressed scenario. 
In case of the scenario represented by BP4, where 
$M_1 < \mu_{eff}$, $\tilde \chi_{2,3}^0$ decays to relatively 
massive of $A_1$.

Remarkably, this signal is observable for some of the BPs corresponding
to comparatively lower masses of $\N0_{2,3}$ and $\C1$ for the 300~$\invfb$ 
luminosity option and very robust for high luminosity option 1000~$\invfb$. 
\begin{figure}[t]
\centering
\includegraphics[width=0.6\textwidth]{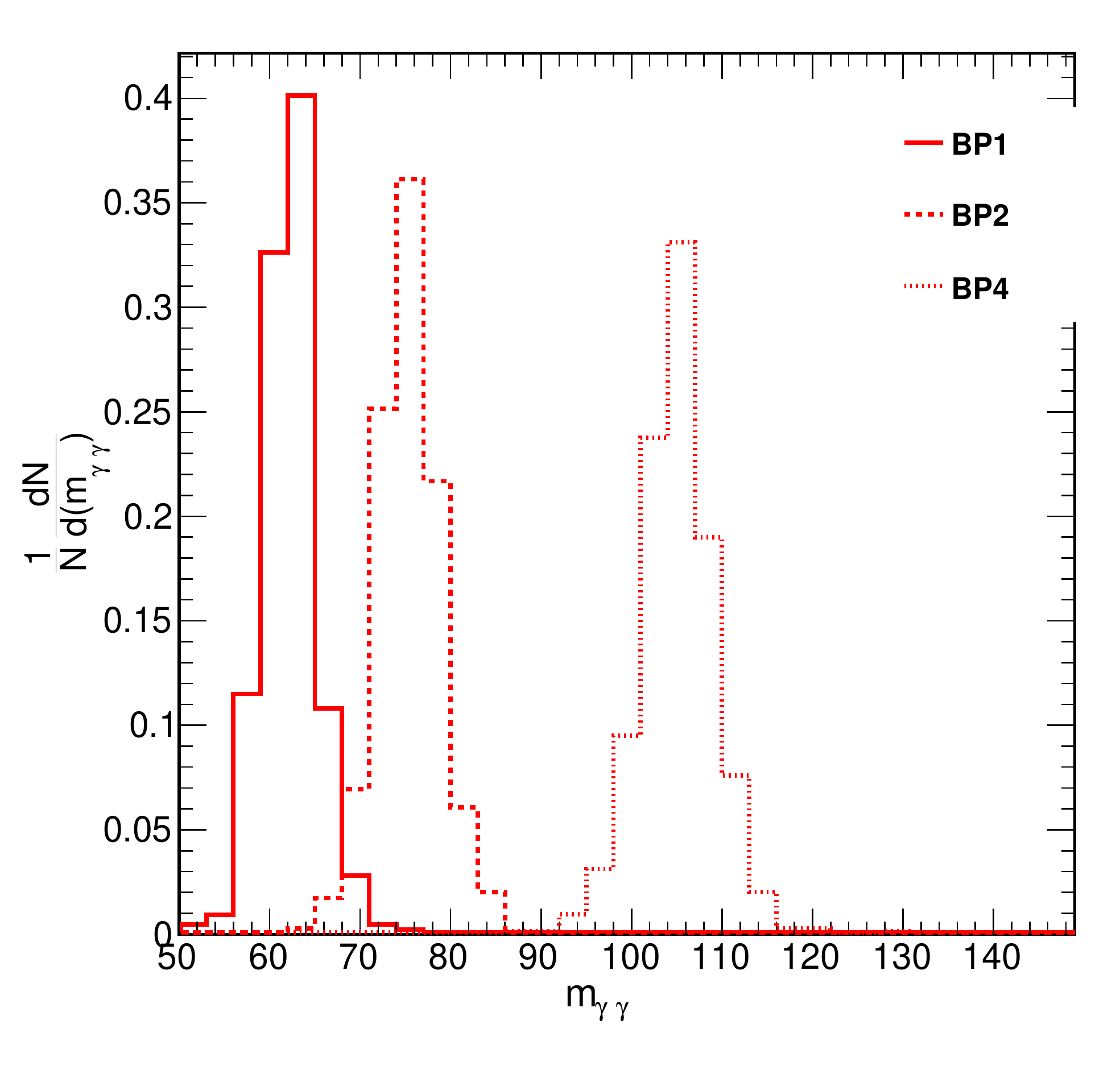}
\caption{Two photon invariant mass for three signal BPs normalizing  
to unity.}
\label{fig:magg}
\end{figure}
Furthermore, it is worth to mention here that in analogy with the 
SM Higgs searches, 
in this study also, the di-photon invariant mass is expected to show a clear 
peak at the mass of $A_1$. In Fig.\ref{fig:magg}, we show the
distribution of reconstructed $m_{\gamma\gamma}$ subject to all cuts 
as listed in 
Table.\ref{tab:event}. Because of the low statistics of background 
events after selection, those are not shown in this figure.
Perhaps, the level of background contamination can be 
reduced further by fitting
the signal peak leading an enhancement of signal sensitivity.
%%%%%%%%%%%%%%%%%%%%%%%%%%%%%%%%%%%%%%%%%%%%%%%%%%%%%%
\section{Summary}
In the NMSSM, one of the non SM-like Higgs boson, 
particularly lightest pseudoscalar $A_1$, which is mostly singlet like, 
can decay to di-photon 
channel via Higgsino like chargino loop with a substantial BR. 
We identify the region of the
parameter space corresponding to BR($\N0_{2,3}\to \N0_1 A_1$) and  
BR$(A_1 \to \gamma\gamma) \ge $, both at the level of 10\%  or more
and present the potential ranges of $\lambda, \kappa $ along with
$\mu_{eff}, A_{\lambda}$. 
We investigate the sensitivity of
the signal $\ell +\gamma\gamma +\slashed E_T$ producing $A_1$ through 
the chargino-neutralino associated production as shown in Eq.\ref{eq:1}. 
The possible 
contamination due to the SM backgrounds are also estimated and 
$W\gamma j$ is found to be the dominant one, where jet fakes as
a photon. Performing a detail simulation of the signal and the 
background processes 
including detector effects using Delphes, we predict the signal sensitivity
for few benchmark points and for a given integrated luminosity options 
for the LHC Run 2 experiments. 
Our simulation shows that this signal is observable marginally 
for 100$\invfb$ integrated
luminosity. However, for larger integrated luminosity option,
this signal is very robust and S/$\sqrt{B} >> 5\sigma$ sensitivity 
can be achieved for the $m_{\N0_{2,3}}, m_{\C1} \sim 400$~GeV 
and $M_{A_1} \sim 70$ GeV, 
where as it severely degrades
for higher masses $\sim$ 600 GeV due to the heavily suppressed cross section.
The reconstructed di-photon invariant mass is expected to show a clear 
visible narrow peak around the mass of $A_1$, which can 
be exploited to suppress backgrounds further to improve  
the signal sensitivity. 
Hence, room for a possible more improvements of signal to background 
ratio exist, which is not explored in the current study. 
We reiterate here that  
two photons BR of Higgs boson is heavily suppressed in the 
SM and as well as in the MSSM. In this context, 
we emphasize again very strongly that this diphoton decay mode 
of $A_1$ can be used as a powerful tool to distinguish the NMSSM 
from the other  SUSY models.
      
\section{Acknowledgment}
JK would like to thank Bibhu P. Mahakud, Jyoti Ranjan Beuria and 
Michael Paraskevas for useful discussions. The authors are also thankful
to Saurabh Nioygi for participating in this project in the beginning.
%%%%%%%%%%%%%%%%%%%%%%%%%%%%%%%%%%%%%%%%%%%%%%%%%%%%%%%%%%%%%%%%%%%%%%%%%%
%%%%%%%%%%%%%%%%%%%%%%%%%%%%%%%%%%%%%%%%%%%%%%%%%%%%%%%%%%%%%%%%%%%%%
\bibliography{paper.bib}

\providecommand{\href}[2]{#2}\begingroup\raggedright\begin{thebibliography}{10}

\bibitem{Aad:2012tfa}
{\bfseries ATLAS} Collaboration, G.~Aad {\em et~al.}, ``{Observation of a new
  particle in the search for the Standard Model Higgs boson with the ATLAS
  detector at the LHC},''
  \href{http://dx.doi.org/10.1016/j.physletb.2012.08.020}{{\em Phys. Lett.}
  {\bfseries B716} (2012) 1--29},
\href{http://arxiv.org/abs/1207.7214}{{\ttfamily arXiv:1207.7214 [hep-ex]}}.
%%CITATION = ARXIV:1207.7214;%%.

\bibitem{Chatrchyan:2012xdj}
{\bfseries CMS} Collaboration, S.~Chatrchyan {\em et~al.}, ``{Observation of a
  new boson at a mass of 125 GeV with the CMS experiment at the LHC},''
  \href{http://dx.doi.org/10.1016/j.physletb.2012.08.021}{{\em Phys. Lett.}
  {\bfseries B716} (2012) 30--61},
\href{http://arxiv.org/abs/1207.7235}{{\ttfamily arXiv:1207.7235 [hep-ex]}}.
%%CITATION = ARXIV:1207.7235;%%.

\bibitem{Hall:2011aa}
L.~J. Hall, D.~Pinner, and J.~T. Ruderman, ``{A Natural SUSY Higgs Near 126
  GeV},'' \href{http://dx.doi.org/10.1007/JHEP04(2012)131}{{\em JHEP}
  {\bfseries 04} (2012) 131},
\href{http://arxiv.org/abs/1112.2703}{{\ttfamily arXiv:1112.2703 [hep-ph]}}.
%%CITATION = ARXIV:1112.2703;%%.

\bibitem{Arbey:2011ab}
A.~Arbey, M.~Battaglia, A.~Djouadi, F.~Mahmoudi, and J.~Quevillon,
  ``{Implications of a 125 GeV Higgs for supersymmetric models},''
  \href{http://dx.doi.org/10.1016/j.physletb.2012.01.053}{{\em Phys. Lett.}
  {\bfseries B708} (2012) 162--169},
\href{http://arxiv.org/abs/1112.3028}{{\ttfamily arXiv:1112.3028 [hep-ph]}}.
%%CITATION = ARXIV:1112.3028;%%.

\bibitem{Kim:1983dt}
J.~E. Kim and H.~P. Nilles, ``{The mu Problem and the Strong CP Problem},''
\href{http://dx.doi.org/10.1016/0370-2693(84)91890-2}{{\em Phys. Lett.}
  {\bfseries B138} (1984) 150--154}.
%%CITATION = PHLTA,B138,150;%%.

\bibitem{Fayet:1974pd}
P.~Fayet, ``{Supergauge Invariant Extension of the Higgs Mechanism and a Model
  for the electron and Its Neutrino},''
\href{http://dx.doi.org/10.1016/0550-3213(75)90636-7}{{\em Nucl. Phys.}
  {\bfseries B90} (1975) 104--124}.
%%CITATION = NUPHA,B90,104;%%.

\bibitem{Ellis:1988er}
J.~R. Ellis, J.~F. Gunion, H.~E. Haber, L.~Roszkowski, and F.~Zwirner, ``{Higgs
  Bosons in a Nonminimal Supersymmetric Model},''
\href{http://dx.doi.org/10.1103/PhysRevD.39.844}{{\em Phys. Rev.} {\bfseries
  D39} (1989) 844}.
%%CITATION = PHRVA,D39,844;%%.

\bibitem{Drees:1988fc}
M.~Drees, ``{Supersymmetric Models with Extended Higgs Sector},''
\href{http://dx.doi.org/10.1142/S0217751X89001448}{{\em Int. J. Mod. Phys.}
  {\bfseries A4} (1989) 3635}.
%%CITATION = IMPAE,A4,3635;%%.

\bibitem{Ellwanger:2009dp}
U.~Ellwanger, C.~Hugonie, and A.~M. Teixeira, ``{The Next-to-Minimal
  Supersymmetric Standard Model},''
  \href{http://dx.doi.org/10.1016/j.physrep.2010.07.001}{{\em Phys. Rept.}
  {\bfseries 496} (2010) 1--77},
\href{http://arxiv.org/abs/0910.1785}{{\ttfamily arXiv:0910.1785 [hep-ph]}}.
%%CITATION = ARXIV:0910.1785;%%.

\bibitem{Miller:2003ay}
D.~J. Miller, R.~Nevzorov, and P.~M. Zerwas, ``{The Higgs sector of the
  next-to-minimal supersymmetric standard model},''
  \href{http://dx.doi.org/10.1016/j.nuclphysb.2003.12.021}{{\em Nucl. Phys.}
  {\bfseries B681} (2004) 3--30},
\href{http://arxiv.org/abs/hep-ph/0304049}{{\ttfamily arXiv:hep-ph/0304049
  [hep-ph]}}.
%%CITATION = HEP-PH/0304049;%%.

\bibitem{Kang:2012sy}
Z.~Kang, J.~Li, and T.~Li, ``{On Naturalness of the MSSM and NMSSM},''
  \href{http://dx.doi.org/10.1007/JHEP11(2012)024}{{\em JHEP} {\bfseries 11}
  (2012) 024},
\href{http://arxiv.org/abs/1201.5305}{{\ttfamily arXiv:1201.5305 [hep-ph]}}.
%%CITATION = ARXIV:1201.5305;%%.

\bibitem{Cao:2013gba}
J.~Cao, F.~Ding, C.~Han, J.~M. Yang, and J.~Zhu, ``{A light Higgs scalar in the
  NMSSM confronted with the latest LHC Higgs data},''
  \href{http://dx.doi.org/10.1007/JHEP11(2013)018}{{\em JHEP} {\bfseries 11}
  (2013) 018},
\href{http://arxiv.org/abs/1309.4939}{{\ttfamily arXiv:1309.4939 [hep-ph]}}.
%%CITATION = ARXIV:1309.4939;%%.

\bibitem{Vasquez:2012hn}
D.~Albornoz~Vasquez, G.~Belanger, C.~Boehm, J.~Da~Silva, P.~Richardson, and
  C.~Wymant, ``{The 125 GeV Higgs in the NMSSM in light of LHC results and
  astrophysics constraints},''
  \href{http://dx.doi.org/10.1103/PhysRevD.86.035023}{{\em Phys. Rev.}
  {\bfseries D86} (2012) 035023},
\href{http://arxiv.org/abs/1203.3446}{{\ttfamily arXiv:1203.3446 [hep-ph]}}.
%%CITATION = ARXIV:1203.3446;%%.

\bibitem{King:2012is}
S.~F. King, M.~Muhlleitner, and R.~Nevzorov, ``{NMSSM Higgs Benchmarks Near 125
  GeV},'' \href{http://dx.doi.org/10.1016/j.nuclphysb.2012.02.010}{{\em Nucl.
  Phys.} {\bfseries B860} (2012) 207--244},
\href{http://arxiv.org/abs/1201.2671}{{\ttfamily arXiv:1201.2671 [hep-ph]}}.
%%CITATION = ARXIV:1201.2671;%%.

\bibitem{Heinemeyer:2011aa}
S.~Heinemeyer, O.~Stal, and G.~Weiglein, ``{Interpreting the LHC Higgs Search
  Results in the MSSM},''
  \href{http://dx.doi.org/10.1016/j.physletb.2012.02.084}{{\em Phys. Lett.}
  {\bfseries B710} (2012) 201--206},
\href{http://arxiv.org/abs/1112.3026}{{\ttfamily arXiv:1112.3026 [hep-ph]}}.
%%CITATION = ARXIV:1112.3026;%%.

\bibitem{Domingo:2015eea}
F.~Domingo and G.~Weiglein, ``{NMSSM interpretations of the observed Higgs
  signal},'' \href{http://dx.doi.org/10.1007/JHEP04(2016)095}{{\em JHEP}
  {\bfseries 04} (2016) 095},
\href{http://arxiv.org/abs/1509.07283}{{\ttfamily arXiv:1509.07283 [hep-ph]}}.
%%CITATION = ARXIV:1509.07283;%%.

\bibitem{Djouadi:2008uw}
A.~Djouadi {\em et~al.}, ``{Benchmark scenarios for the NMSSM},''
  \href{http://dx.doi.org/10.1088/1126-6708/2008/07/002}{{\em JHEP} {\bfseries
  07} (2008) 002},
\href{http://arxiv.org/abs/0801.4321}{{\ttfamily arXiv:0801.4321 [hep-ph]}}.
%%CITATION = ARXIV:0801.4321;%%.

\bibitem{King:2012tr}
S.~F. King, M.~Mühlleitner, R.~Nevzorov, and K.~Walz, ``{Natural NMSSM Higgs
  Bosons},'' \href{http://dx.doi.org/10.1016/j.nuclphysb.2013.01.020}{{\em
  Nucl. Phys.} {\bfseries B870} (2013) 323--352},
\href{http://arxiv.org/abs/1211.5074}{{\ttfamily arXiv:1211.5074 [hep-ph]}}.
%%CITATION = ARXIV:1211.5074;%%.

\bibitem{Christensen:2013dra}
N.~D. Christensen, T.~Han, Z.~Liu, and S.~Su, ``{Low-Mass Higgs Bosons in the
  NMSSM and Their LHC Implications},''
  \href{http://dx.doi.org/10.1007/JHEP08(2013)019}{{\em JHEP} {\bfseries 08}
  (2013) 019},
\href{http://arxiv.org/abs/1303.2113}{{\ttfamily arXiv:1303.2113 [hep-ph]}}.
%%CITATION = ARXIV:1303.2113;%%.

\bibitem{Kumar:2016vhm}
J.~Kumar and M.~Paraskevas, ``{Distinguishing between MSSM and NMSSM through
  $\Delta F=2$ processes},''
\href{http://arxiv.org/abs/1608.08794}{{\ttfamily arXiv:1608.08794 [hep-ph]}}.
%%CITATION = ARXIV:1608.08794;%%.

\bibitem{Guchait:2015owa}
M.~Guchait and J.~Kumar, ``{Light Higgs Bosons in NMSSM at the LHC},''
  \href{http://dx.doi.org/10.1142/S0217751X1650069X}{{\em Int. J. Mod. Phys.}
  {\bfseries A31} no.~12, (2016) 1650069},
\href{http://arxiv.org/abs/1509.02452}{{\ttfamily arXiv:1509.02452 [hep-ph]}}.
%%CITATION = ARXIV:1509.02452;%%.

\bibitem{Gunion:2012zd}
J.~F. Gunion, Y.~Jiang, and S.~Kraml, ``{The Constrained NMSSM and Higgs near
  125 GeV},'' \href{http://dx.doi.org/10.1016/j.physletb.2012.03.027}{{\em
  Phys. Lett.} {\bfseries B710} (2012) 454--459},
\href{http://arxiv.org/abs/1201.0982}{{\ttfamily arXiv:1201.0982 [hep-ph]}}.
%%CITATION = ARXIV:1201.0982;%%.

\bibitem{Ellwanger:2012ke}
U.~Ellwanger and C.~Hugonie, ``{Higgs bosons near 125 GeV in the NMSSM with
  constraints at the GUT scale},''
  \href{http://dx.doi.org/10.1155/2012/625389}{{\em Adv. High Energy Phys.}
  {\bfseries 2012} (2012) 625389},
\href{http://arxiv.org/abs/1203.5048}{{\ttfamily arXiv:1203.5048 [hep-ph]}}.
%%CITATION = ARXIV:1203.5048;%%.

\bibitem{Badziak:2013bda}
M.~Badziak, M.~Olechowski, and S.~Pokorski, ``{New Regions in the NMSSM with a
  125 GeV Higgs},'' \href{http://dx.doi.org/10.1007/JHEP06(2013)043}{{\em JHEP}
  {\bfseries 06} (2013) 043},
\href{http://arxiv.org/abs/1304.5437}{{\ttfamily arXiv:1304.5437 [hep-ph]}}.
%%CITATION = ARXIV:1304.5437;%%.

\bibitem{Chatrchyan:2012am}
{\bfseries CMS} Collaboration, S.~Chatrchyan {\em et~al.}, ``{Search for a
  light pseudoscalar Higgs boson in the dimuon decay channel in $pp$ collisions
  at $\sqrt{s}=7$ TeV},''
  \href{http://dx.doi.org/10.1103/PhysRevLett.109.121801}{{\em Phys. Rev.
  Lett.} {\bfseries 109} (2012) 121801},
\href{http://arxiv.org/abs/1206.6326}{{\ttfamily arXiv:1206.6326 [hep-ex]}}.
%%CITATION = ARXIV:1206.6326;%%.

\bibitem{Khachatryan:2015baw}
{\bfseries CMS} Collaboration, V.~Khachatryan {\em et~al.}, ``{Search for a
  low-mass pseudoscalar Higgs boson produced in association with a $b\bar{b}$
  pair in pp collisions at $\sqrt{s} =$ 8 TeV},''
  \href{http://dx.doi.org/10.1016/j.physletb.2016.05.003}{{\em Phys. Lett.}
  {\bfseries B758} (2016) 296--320},
\href{http://arxiv.org/abs/1511.03610}{{\ttfamily arXiv:1511.03610 [hep-ex]}}.
%%CITATION = ARXIV:1511.03610;%%.

\bibitem{Aad:2015oqa}
{\bfseries ATLAS} Collaboration, G.~Aad {\em et~al.}, ``{Search for Higgs
  bosons decaying to $aa$ in the $\mu\mu\tau\tau$ final state in $pp$
  collisions at $\sqrt{s} = $ 8 TeV with the ATLAS experiment},''
  \href{http://dx.doi.org/10.1103/PhysRevD.92.052002}{{\em Phys. Rev.}
  {\bfseries D92} no.~5, (2015) 052002},
\href{http://arxiv.org/abs/1505.01609}{{\ttfamily arXiv:1505.01609 [hep-ex]}}.
%%CITATION = ARXIV:1505.01609;%%.

\bibitem{Aad:2015bua}
{\bfseries ATLAS} Collaboration, G.~Aad {\em et~al.}, ``{Search for new
  phenomena in events with at least three photons collected in $pp$ collisions
  at $\sqrt{s}$ = 8 TeV with the ATLAS detector},''
  \href{http://dx.doi.org/10.1140/epjc/s10052-016-4034-8}{{\em Eur. Phys. J.}
  {\bfseries C76} no.~4, (2016) 210},
\href{http://arxiv.org/abs/1509.05051}{{\ttfamily arXiv:1509.05051 [hep-ex]}}.
%%CITATION = ARXIV:1509.05051;%%.

\bibitem{Ellwanger:2004gz}
U.~Ellwanger, J.~F. Gunion, C.~Hugonie, and S.~Moretti, ``{NMSSM Higgs
  discovery at the LHC},'' in {\em {Physics at TeV colliders. Proceedings,
  Workshop, Les Houches, France, May 26-June 3, 2003}}.
\newblock 2004.
\newblock
\href{http://arxiv.org/abs/hep-ph/0401228}{{\ttfamily arXiv:hep-ph/0401228
  [hep-ph]}}.
\newblock
%%CITATION = HEP-PH/0401228;%%.

\bibitem{Ellwanger:2011sk}
U.~Ellwanger, ``{Higgs Bosons in the Next-to-Minimal Supersymmetric Standard
  Model at the LHC},''
  \href{http://dx.doi.org/10.1140/epjc/s10052-011-1782-3}{{\em Eur. Phys. J.}
  {\bfseries C71} (2011) 1782},
\href{http://arxiv.org/abs/1108.0157}{{\ttfamily arXiv:1108.0157 [hep-ph]}}.
%%CITATION = ARXIV:1108.0157;%%.

\bibitem{King:2014xwa}
S.~F. King, M.~Mühlleitner, R.~Nevzorov, and K.~Walz, ``{Discovery Prospects
  for NMSSM Higgs Bosons at the High-Energy Large Hadron Collider},''
  \href{http://dx.doi.org/10.1103/PhysRevD.90.095014}{{\em Phys. Rev.}
  {\bfseries D90} no.~9, (2014) 095014},
\href{http://arxiv.org/abs/1408.1120}{{\ttfamily arXiv:1408.1120 [hep-ph]}}.
%%CITATION = ARXIV:1408.1120;%%.

\bibitem{Mahmoudi:2010xp}
F.~Mahmoudi, J.~Rathsman, O.~Stal, and L.~Zeune, ``{Light Higgs bosons in
  phenomenological NMSSM},''
  \href{http://dx.doi.org/10.1140/epjc/s10052-011-1608-3}{{\em Eur. Phys. J.}
  {\bfseries C71} (2011) 1608},
\href{http://arxiv.org/abs/1012.4490}{{\ttfamily arXiv:1012.4490 [hep-ph]}}.
%%CITATION = ARXIV:1012.4490;%%.

\bibitem{Bomark:2014gya}
N.-E. Bomark, S.~Moretti, S.~Munir, and L.~Roszkowski, ``{A light NMSSM
  pseudoscalar Higgs boson at the LHC redux},''
  \href{http://dx.doi.org/10.1007/JHEP02(2015)044}{{\em JHEP} {\bfseries 02}
  (2015) 044},
\href{http://arxiv.org/abs/1409.8393}{{\ttfamily arXiv:1409.8393 [hep-ph]}}.
%%CITATION = ARXIV:1409.8393;%%.

\bibitem{Belyaev:2008gj}
A.~Belyaev, S.~Hesselbach, S.~Lehti, S.~Moretti, A.~Nikitenko, and C.~H.
  Shepherd-Themistocleous, ``{The Scope of the 4 tau Channel in Higgs-strahlung
  and Vector Boson Fusion for the NMSSM No-Lose Theorem at the LHC},''
\href{http://arxiv.org/abs/0805.3505}{{\ttfamily arXiv:0805.3505 [hep-ph]}}.
%%CITATION = ARXIV:0805.3505;%%.

\bibitem{Belyaev:2010ka}
A.~Belyaev, J.~Pivarski, A.~Safonov, S.~Senkin, and A.~Tatarinov, ``{LHC
  discovery potential of the lightest NMSSM Higgs in the h1 -> a1 a1 -> 4 muons
  channel},'' \href{http://dx.doi.org/10.1103/PhysRevD.81.075021}{{\em Phys.
  Rev.} {\bfseries D81} (2010) 075021},
\href{http://arxiv.org/abs/1002.1956}{{\ttfamily arXiv:1002.1956 [hep-ph]}}.
%%CITATION = ARXIV:1002.1956;%%.

\bibitem{Almarashi:2011hj}
M.~M. Almarashi and S.~Moretti, ``{Muon Signals of Very Light CP-odd Higgs
  states of the NMSSM at the LHC},''
  \href{http://dx.doi.org/10.1103/PhysRevD.83.035023}{{\em Phys. Rev.}
  {\bfseries D83} (2011) 035023},
\href{http://arxiv.org/abs/1101.1137}{{\ttfamily arXiv:1101.1137 [hep-ph]}}.
%%CITATION = ARXIV:1101.1137;%%.

\bibitem{Cerdeno:2013cz}
D.~G. Cerdeno, P.~Ghosh, and C.~B. Park, ``{Probing the two light Higgs
  scenario in the NMSSM with a low-mass pseudoscalar},''
  \href{http://dx.doi.org/10.1007/JHEP06(2013)031}{{\em JHEP} {\bfseries 06}
  (2013) 031},
\href{http://arxiv.org/abs/1301.1325}{{\ttfamily arXiv:1301.1325 [hep-ph]}}.
%%CITATION = ARXIV:1301.1325;%%.

\bibitem{Curtin:2014pda}
D.~Curtin, R.~Essig, and Y.-M. Zhong, ``{Uncovering light scalars with exotic
  Higgs decays to $ b\overline{b}{\mu}^{+}{\mu}^{-} $},''
  \href{http://dx.doi.org/10.1007/JHEP06(2015)025}{{\em JHEP} {\bfseries 06}
  (2015) 025},
\href{http://arxiv.org/abs/1412.4779}{{\ttfamily arXiv:1412.4779 [hep-ph]}}.
%%CITATION = ARXIV:1412.4779;%%.

\bibitem{Bomark:2015fga}
N.-E. Bomark, S.~Moretti, and L.~Roszkowski, ``{Detection prospects of light
  NMSSM Higgs pseudoscalar via cascades of heavier scalars from vector boson
  fusion and Higgs-strahlung},''
  \href{http://dx.doi.org/10.1088/0954-3899/43/10/105003}{{\em J. Phys.}
  {\bfseries G43} no.~10, (2016) 105003},
\href{http://arxiv.org/abs/1503.04228}{{\ttfamily arXiv:1503.04228 [hep-ph]}}.
%%CITATION = ARXIV:1503.04228;%%.

\bibitem{Arhrib:2006sx}
A.~Arhrib, K.~Cheung, T.-J. Hou, and K.-W. Song, ``{Associated production of a
  light pseudoscalar Higgs boson with a chargino pair in the NMSSM},''
  \href{http://dx.doi.org/10.1088/1126-6708/2007/03/073}{{\em JHEP} {\bfseries
  03} (2007) 073},
\href{http://arxiv.org/abs/hep-ph/0606114}{{\ttfamily arXiv:hep-ph/0606114
  [hep-ph]}}.
%%CITATION = HEP-PH/0606114;%%.

\bibitem{Dermisek:2007yt}
R.~Dermisek and J.~F. Gunion, ``{The NMSSM Solution to the Fine-Tuning Problem,
  Precision Electroweak Constraints and the Largest LEP Higgs Event Excess},''
  \href{http://dx.doi.org/10.1103/PhysRevD.76.095006}{{\em Phys. Rev.}
  {\bfseries D76} (2007) 095006},
\href{http://arxiv.org/abs/0705.4387}{{\ttfamily arXiv:0705.4387 [hep-ph]}}.
%%CITATION = ARXIV:0705.4387;%%.

\bibitem{Kim:2012az}
J.~E. Kim, H.~P. Nilles, and M.-S. Seo, ``{Singlet Superfield Extension of the
  Minimal Supersymmetric Standard model with Peccei-Quinn symmetry and a Light
  Pseudoscalar Higgs Boson at the LHC},''
  \href{http://dx.doi.org/10.1142/S0217732312501660}{{\em Mod. Phys. Lett.}
  {\bfseries A27} (2012) 1250166},
\href{http://arxiv.org/abs/1201.6547}{{\ttfamily arXiv:1201.6547 [hep-ph]}}.
%%CITATION = ARXIV:1201.6547;%%.

\bibitem{Ellwanger:2015uaz}
U.~Ellwanger and M.~Rodriguez-Vazquez, ``{Discovery Prospects of a Light Scalar
  in the NMSSM},'' \href{http://dx.doi.org/10.1007/JHEP02(2016)096}{{\em JHEP}
  {\bfseries 02} (2016) 096},
\href{http://arxiv.org/abs/1512.04281}{{\ttfamily arXiv:1512.04281 [hep-ph]}}.
%%CITATION = ARXIV:1512.04281;%%.

\bibitem{Moretti:2006sv}
S.~Moretti and S.~Munir, ``{Di-photon Higgs signals at the LHC in the
  next-to-minimal supersymmetric standard model},''
  \href{http://dx.doi.org/10.1140/epjc/s2006-02585-7}{{\em Eur. Phys. J.}
  {\bfseries C47} (2006) 791--803},
\href{http://arxiv.org/abs/hep-ph/0603085}{{\ttfamily arXiv:hep-ph/0603085
  [hep-ph]}}.
%%CITATION = HEP-PH/0603085;%%.

\bibitem{Ghosh:2012mc}
D.~Ghosh, M.~Guchait, and D.~Sengupta, ``{Higgs Signal in Chargino-Neutralino
  Production at the LHC},''
  \href{http://dx.doi.org/10.1140/epjc/s10052-012-2141-8}{{\em Eur. Phys. J.}
  {\bfseries C72} (2012) 2141},
\href{http://arxiv.org/abs/1202.4937}{{\ttfamily arXiv:1202.4937 [hep-ph]}}.
%%CITATION = ARXIV:1202.4937;%%.

\bibitem{Cerdeno:2013qta}
D.~G. Cerdeño, P.~Ghosh, C.~B. Park, and M.~Peiró, ``{Collider signatures of
  a light NMSSM pseudoscalar in neutralino decays in the light of LHC
  results},'' \href{http://dx.doi.org/10.1007/JHEP02(2014)048}{{\em JHEP}
  {\bfseries 02} (2014) 048},
\href{http://arxiv.org/abs/1307.7601}{{\ttfamily arXiv:1307.7601 [hep-ph]}}.
%%CITATION = ARXIV:1307.7601;%%.

\bibitem{Guchait:1991ia}
M.~Guchait, ``{Exact solution of the neutralino mass matrix},''
  \href{http://dx.doi.org/10.1007/BF01555748}{{\em Z. Phys.} {\bfseries C57}
  (1993) 157--164}.
[Erratum: Z. Phys.C61,178(1994)].
%%CITATION = ZEPYA,C57,157;%%.

\bibitem{Choi:2001ww}
S.~Y. Choi, J.~Kalinowski, G.~A. Moortgat-Pick, and P.~M. Zerwas, ``{Analysis
  of the neutralino system in supersymmetric theories},''
  \href{http://dx.doi.org/10.1007/s100520100808}{{\em Eur. Phys. J.} {\bfseries
  C22} (2001) 563--579}, \href{http://arxiv.org/abs/hep-ph/0108117}{{\ttfamily
  arXiv:hep-ph/0108117 [hep-ph]}}.
[Addendum: Eur. Phys. J.C23,769(2002)].
%%CITATION = HEP-PH/0108117;%%.

\bibitem{Pandita:1994vw}
P.~N. Pandita, ``{Neutralino mass matrix in the nonminimal supersymmetric
  Standard model},''
\href{http://dx.doi.org/10.1007/BF01557633}{{\em Z. Phys.} {\bfseries C63}
  (1994) 659--671}.
%%CITATION = ZEPYA,C63,659;%%.

\bibitem{Choi:2004zx}
S.~Y. Choi, D.~J. Miller, and P.~M. Zerwas, ``{The Neutralino sector of the
  next-to-minimal supersymmetric standard model},''
  \href{http://dx.doi.org/10.1016/j.nuclphysb.2005.01.006}{{\em Nucl. Phys.}
  {\bfseries B711} (2005) 83--111},
\href{http://arxiv.org/abs/hep-ph/0407209}{{\ttfamily arXiv:hep-ph/0407209
  [hep-ph]}}.
%%CITATION = HEP-PH/0407209;%%.

\bibitem{Beenakker:1999xh}
W.~Beenakker, M.~Klasen, M.~Kramer, T.~Plehn, M.~Spira, and P.~M. Zerwas,
  ``{The Production of charginos / neutralinos and sleptons at hadron
  colliders},'' \href{http://dx.doi.org/10.1103/PhysRevLett.100.029901,
  10.1103/PhysRevLett.83.3780}{{\em Phys. Rev. Lett.} {\bfseries 83} (1999)
  3780--3783}, \href{http://arxiv.org/abs/hep-ph/9906298}{{\ttfamily
  arXiv:hep-ph/9906298 [hep-ph]}}.
[Erratum: Phys. Rev. Lett.100,029901(2008)].
%%CITATION = HEP-PH/9906298;%%.

\bibitem{Lai:2010vv}
H.-L. Lai, M.~Guzzi, J.~Huston, Z.~Li, P.~M. Nadolsky, J.~Pumplin, and C.~P.
  Yuan, ``{New parton distributions for collider physics},''
  \href{http://dx.doi.org/10.1103/PhysRevD.82.074024}{{\em Phys. Rev.}
  {\bfseries D82} (2010) 074024},
\href{http://arxiv.org/abs/1007.2241}{{\ttfamily arXiv:1007.2241 [hep-ph]}}.
%%CITATION = ARXIV:1007.2241;%%.

\bibitem{Beenakker:1996ed}
W.~Beenakker, R.~Hopker, and M.~Spira, ``{PROSPINO: A Program for the
  production of supersymmetric particles in next-to-leading order QCD},''
\href{http://arxiv.org/abs/hep-ph/9611232}{{\ttfamily arXiv:hep-ph/9611232
  [hep-ph]}}.
%%CITATION = HEP-PH/9611232;%%.

\bibitem{Ellwanger:2004xm}
U.~Ellwanger, J.~F. Gunion, and C.~Hugonie, ``{NMHDECAY: A Fortran code for the
  Higgs masses, couplings and decay widths in the NMSSM},''
  \href{http://dx.doi.org/10.1088/1126-6708/2005/02/066}{{\em JHEP} {\bfseries
  02} (2005) 066},
\href{http://arxiv.org/abs/hep-ph/0406215}{{\ttfamily arXiv:hep-ph/0406215
  [hep-ph]}}.
%%CITATION = HEP-PH/0406215;%%.

\bibitem{Munir:2013wka}
S.~Munir, L.~Roszkowski, and S.~Trojanowski, ``{Simultaneous enhancement in
  $\gamma \gamma, b\bar{b}$ and $\tau^{+} \tau^{-}$ rates in the NMSSM with
  nearly degenerate scalar and pseudoscalar Higgs bosons},''
  \href{http://dx.doi.org/10.1103/PhysRevD.88.055017}{{\em Phys. Rev.}
  {\bfseries D88} no.~5, (2013) 055017},
\href{http://arxiv.org/abs/1305.0591}{{\ttfamily arXiv:1305.0591 [hep-ph]}}.
%%CITATION = ARXIV:1305.0591;%%.

\bibitem{Spira:1995rr}
M.~Spira, A.~Djouadi, D.~Graudenz, and P.~M. Zerwas, ``{Higgs boson production
  at the LHC},'' \href{http://dx.doi.org/10.1016/0550-3213(95)00379-7}{{\em
  Nucl. Phys.} {\bfseries B453} (1995) 17--82},
\href{http://arxiv.org/abs/hep-ph/9504378}{{\ttfamily arXiv:hep-ph/9504378
  [hep-ph]}}.
%%CITATION = HEP-PH/9504378;%%.

\bibitem{Spira:1997dg}
M.~Spira, ``{QCD effects in Higgs physics},''
  \href{http://dx.doi.org/10.1002/(SICI)1521-3978(199804)46:3<203::AID-PROP203>3.0.CO;2-4}{{\em
  Fortsch. Phys.} {\bfseries 46} (1998) 203--284},
\href{http://arxiv.org/abs/hep-ph/9705337}{{\ttfamily arXiv:hep-ph/9705337
  [hep-ph]}}.
%%CITATION = HEP-PH/9705337;%%.

\bibitem{Khachatryan:2016vau}
{\bfseries ATLAS, CMS} Collaboration, G.~Aad {\em et~al.}, ``{Measurements of
  the Higgs boson production and decay rates and constraints on its couplings
  from a combined ATLAS and CMS analysis of the LHC $pp$ collision data at
  $\sqrt{s}=$ 7 and 8 TeV},''
\href{http://arxiv.org/abs/1606.02266}{{\ttfamily arXiv:1606.02266 [hep-ex]}}.
%%CITATION = ARXIV:1606.02266;%%.

\bibitem{Sjostrand:2006za}
T.~Sjostrand, S.~Mrenna, and P.~Z. Skands, ``{PYTHIA 6.4 Physics and Manual},''
  \href{http://dx.doi.org/10.1088/1126-6708/2006/05/026}{{\em JHEP} {\bfseries
  05} (2006) 026},
\href{http://arxiv.org/abs/hep-ph/0603175}{{\ttfamily arXiv:hep-ph/0603175
  [hep-ph]}}.
%%CITATION = HEP-PH/0603175;%%.

\bibitem{Skands:2003cj}
P.~Z. Skands {\em et~al.}, ``{SUSY Les Houches accord: Interfacing SUSY
  spectrum calculators, decay packages, and event generators},''
  \href{http://dx.doi.org/10.1088/1126-6708/2004/07/036}{{\em JHEP} {\bfseries
  07} (2004) 036},
\href{http://arxiv.org/abs/hep-ph/0311123}{{\ttfamily arXiv:hep-ph/0311123
  [hep-ph]}}.
%%CITATION = HEP-PH/0311123;%%.

\bibitem{Aad:2015jqa}
{\bfseries ATLAS} Collaboration, G.~Aad {\em et~al.}, ``{Search for direct pair
  production of a chargino and a neutralino decaying to the 125 GeV Higgs boson
  in $\sqrt{s} = 8$ TeV ${pp}$ collisions with the ATLAS detector},''
  \href{http://dx.doi.org/10.1140/epjc/s10052-015-3408-7}{{\em Eur. Phys. J.}
  {\bfseries C75} no.~5, (2015) 208},
\href{http://arxiv.org/abs/1501.07110}{{\ttfamily arXiv:1501.07110 [hep-ex]}}.
%%CITATION = ARXIV:1501.07110;%%.

\bibitem{stdhep}
P.~L. L.~Garren, ``{ StdHep User Manual},''.

\bibitem{deFavereau:2013fsa}
{\bfseries DELPHES 3} Collaboration, J.~de~Favereau, C.~Delaere, P.~Demin,
  A.~Giammanco, V.~Lemaître, A.~Mertens, and M.~Selvaggi, ``{DELPHES 3, A
  modular framework for fast simulation of a generic collider experiment},''
  \href{http://dx.doi.org/10.1007/JHEP02(2014)057}{{\em JHEP} {\bfseries 02}
  (2014) 057},
\href{http://arxiv.org/abs/1307.6346}{{\ttfamily arXiv:1307.6346 [hep-ex]}}.
%%CITATION = ARXIV:1307.6346;%%.

\bibitem{Alwall:2014hca}
J.~Alwall, R.~Frederix, S.~Frixione, V.~Hirschi, F.~Maltoni, O.~Mattelaer,
  H.~S. Shao, T.~Stelzer, P.~Torrielli, and M.~Zaro, ``{The automated
  computation of tree-level and next-to-leading order differential cross
  sections, and their matching to parton shower simulations},''
  \href{http://dx.doi.org/10.1007/JHEP07(2014)079}{{\em JHEP} {\bfseries 07}
  (2014) 079},
\href{http://arxiv.org/abs/1405.0301}{{\ttfamily arXiv:1405.0301 [hep-ph]}}.
%%CITATION = ARXIV:1405.0301;%%.

\end{thebibliography}\endgroup
\bibliographystyle{utphys.bst}
\end{document}